\documentclass[footinbib,fleqn,aps,prb,twocolumn,superscriptaddress,reprint]{revtex4-2}
\usepackage{xr}
\usepackage[]{amsmath}
\usepackage{amssymb}
\usepackage[dvips]{graphicx}
\usepackage{color}
\usepackage{tabularx}
\usepackage{mathtools}
\usepackage{algpseudocode}
\usepackage{enumitem}
\usepackage{multirow}
\usepackage{epstopdf}  
\usepackage{adjustbox}
\usepackage{bm}
\usepackage{float}
\usepackage{siunitx}
\usepackage{chemformula}

\usepackage{makecell}
\usepackage{hyperref}
\usepackage[version=3]{mhchem}
\usepackage{tikz}
\usetikzlibrary{shapes,snakes}
\usetikzlibrary{arrows}
\usepackage{threeparttable}
\usepackage{natbib}

\newcommand{\Tc}{\textit{T}$_c$}

\begin{document}

\title{Prediction and Synthesis of \ch{Mg4Pt3H6}: A Metallic Complex Transition Metal Hydride Stabilized at Ambient Pressure}
  
\author{Wencheng Lu}
\email{wlu@carnegiescience.edu}
\affiliation{Earth and Planets Laboratory, Carnegie Institution for Science, 5241 Broad Branch Road NW, Washington, DC 20015, USA}

\author{Michael J. Hutcheon}
\affiliation{Enterprise Science Fund, Intellectual Ventures, 3150 139th Ave SE, Bellevue, WA, 98005, USA}

\author{Mads F. Hansen}
\affiliation{Earth and Planets Laboratory, Carnegie Institution for Science, 5241 Broad Branch Road NW, Washington, DC 20015, USA}
\affiliation{Cavendish Laboratory, University of Cambridge, JJ Thomson Avenue,
Cambridge, CB3 0HE, United Kingdom}

\author{Kapildeb Dolui}
\affiliation{Department of Materials Science and Metallurgy, University of Cambridge, 27 Charles Babbage Road, Cambridge, CB3 0FS, UK}

\author{Shubham Sinha}
\affiliation{Earth and Planets Laboratory, Carnegie Institution for Science, 5241 Broad Branch Road NW, Washington, DC 20015, USA}

\author{Mihir R. Sahoo}
\affiliation{Institute of Theoretical and Computational Physics, Graz University of Technology, NAWI Graz, 8010 Graz, Austria}

\author{Chris J. Pickard}
\affiliation{Department of Materials Science and Metallurgy, University of Cambridge, 27 Charles Babbage Road, Cambridge, CB3 0FS, UK}
\affiliation{Advanced Institute for Materials Research, Tohoku University, Sendai, 980-8577, Japan}

\author{Christoph Heil}
\affiliation{Institute of Theoretical and Computational Physics, Graz University of Technology, NAWI Graz, 8010 Graz, Austria}

\author{Anna Pakhomova}
\affiliation{European Synchrotron Radiation Facility, B.P.220, F-38043 Grenoble Cedex, France}

\author{Mohamed Mezouar}
\affiliation{European Synchrotron Radiation Facility, B.P.220, F-38043 Grenoble Cedex, France}

\author{Dominik Daisenberger}
\affiliation{Diamond Light Source, Chilton, Didcot OX11 0DE, United Kingdom}

\author{Stella Chariton}
\affiliation{Center for Advanced Radiation Sources, The University of Chicago, Lemont, Illinois 60439, USA}

\author{Vitali Prakapenka}
\affiliation{Center for Advanced Radiation Sources, The University of Chicago, Lemont, Illinois 60439, USA}

\author{Matthew N. Julian}
\affiliation{Enterprise Science Fund, Intellectual Ventures, 3150 139th Ave SE, Bellevue, WA, 98005, USA}

\author{Rohit P. Prasankumar}
\affiliation{Enterprise Science Fund, Intellectual Ventures, 3150 139th Ave SE, Bellevue, WA, 98005, USA}

\author{Timothy A. Strobel}
\email{tstrobel@carnegiescience.edu}
\affiliation{Earth and Planets Laboratory, Carnegie Institution for Science, 5241 Broad Branch Road NW, Washington, DC 20015, USA}

\date{\today}
	
\begin{abstract}
The low-pressure stabilization of superconducting hydrides with high critical temperatures (\Tc s) remains a significant challenge, and experimentally verified superconducting hydrides are generally constrained to a limited number of structural prototypes. Ternary transition-metal complex hydrides (hydrido complexes)---typically regarded as hydrogen storage materials---exhibit a large range of compounds stabilized at low pressure with recent predictions for high-\Tc\ superconductivity. Motivated by this class of materials, we investigated complex hydride formation in the Mg--Pt--H system, which has no known ternary hydride compounds. Guided by \textit{ab initio} structural predictions, we successfully synthesized a novel complex transition-metal hydride, \ch{Mg4Pt3H6}, using laser-heated diamond anvil cells. The compound forms in a body-centered cubic structural prototype at moderate pressures between $\sim$8--\SI{25}{GPa}. Unlike the majority of known hydrido complexes, \ch{Mg4Pt3H6} is metallic, with formal charge described as 4[Mg]$^{2+}\cdot$~3[PtH$_2$]$^{2-}$. X-ray diffraction (XRD) measurements obtained during decompression reveal that \ch{Mg4Pt3H6} remains stable upon quenching to ambient conditions. Magnetic-field and temperature-dependent electrical transport measurements indicate ambient-pressure superconductivity with \Tc\ (50\%) = \SI{2.9}{K}, in reasonable agreement with theoretical calculations. These findings clarify the phase behavior in the Mg--Pt--H system and provide valuable insights for transition-metal complex hydrides as a new class of hydrogen-rich superconductors. 
\end{abstract}

\maketitle

\section{Introduction}

Hydrides hold significant scientific and practical importance as candidates for high-temperature superconductors and energy storage materials. In particular, the remarkable superconducting critical temperatures (\textit{T}$_c$s) observed in hydride materials have drawn intense research interest in recent years~\cite{sun2024clathrate,gao2021superconducting}. Many studies indicate high-\Tc\ superconductivity in binary hydrides at pressures exceeding \SI{150}{GPa}, including H$_3$S~\cite{drozdov2015conventional}, CaH$_6$~\cite{ma2022high,wang2012superconductive,li2022superconductivity}, YH$_6$ \cite{kong2021superconductivity,troyan2021anomalous,peng2017hydrogen}, YH$_9$~\cite{peng2017hydrogen,kong2021superconductivity,snider2021synthesis}, and LaH$_{10}$~\cite{peng2017hydrogen,liu2017potential,drozdov2019superconductivity,somayazulu2019evidence}. Nevertheless, the low-pressure recovery/stabilization of these materials or discovery of alternative high-\Tc\ hydrides at near-ambient pressure remains a major challenge.

The configurational space of binary hydrides is relatively constrained, with limited structural and compositional variability in known high-\Tc\ compounds. Consequently, attention has shifted towards ternary hydrides, which offer larger structural diversity and a wider range of possible prototypes. Ternary hydrides offer the potential for enhanced physical and chemical properties through the tuning of additional degrees of freedom, and may enable enhanced performance and versatility~\cite{zhang2022superconducting}. An intuitive approach to generating ternary hydrides involves the substitution of elements within the metal sublattice, as demonstrated by alloy compounds like \ch{(La,Y)H6}~\cite{semenok2021superconductivity}, \ch{(La,Y)H10}~\cite{semenok2021superconductivity}, \ch{(La,Ca)H10}~\cite{chen2024synthesis}, \ch{(La,Al)H10}~\cite{chen2024high}, \ch{(La,Ce)H9}~\cite{bi2022giant}, \ch{(Y,Ce)H9}~\cite{chen2024synthesis}, and \ch{(Lu,Y)4H23}~\cite{zhang2025synthesis}. This type of alloying approach typically involves trade-offs between pressure stability and superconducting transition temperatures. To date, this approach has not significantly expanded the richness of ternary prototype structures. The search for stoichiometric ternary compounds with well-defined crystal structures remains a significant and ongoing challenge. Extensive efforts have been devoted to theoretically predicting superconductors in new ternary hydrides~\cite{du2024high,chen2024high,jiang2024ternary,he2024predicted,lucrezi2023quantum,zhang2022design,Zhao2022emergent,gao2021phonon,sun2019route}. However, only a few of these predictions have been experimentally confirmed. Recently, a stoichiometric ternary hydride in a fluorite-type backbone, \ch{LaBeH8}, was synthesized with \Tc\ = \SI{110}{K} at \SI{80}{GPa}~\cite{song2023stoichiometric}. Another lanthanum-based borohydride \ch{LaB2H8} was experimentally reported with \Tc\ = \SI{106}{K} at \SI{90} {GPa}~\cite{song2024superconductivity}.

Notably, a well-defined class of ternary hydrides, known as transition-metal complex hydrides~\cite{Yvon2006}, offers a broad range of structural and chemical diversity. These complex hydrides---typically containing transition metals from groups VII (Mn) to XII (Zn)---consist of alkali or alkaline-earth metal cations paired with homoleptic transition-metal-hydride complex anions (hydrido complexes), and are regarded as promising candidates for hydrogen storage. Their electronic configurations are typically analogous to those of coordination compounds, following conventional electron-counting rules through charge transfer within the surrounding cationic matrix~\cite{Yvon2006}. The geometric configurations of hydrido complexes can be categorized into distinct types based on the coordination number of the central transition metal. Examples include tricapped trigonal prismatic (CN=9)~\cite{bronger1999k2reh9}, pentagonal bipyramidal (CN=7)~\cite{bronger2002new}, octahedral (CN=6)~\cite{orgaz2000electronic}, square pyramidal (CN=5)~\cite{ivanov1989ternary}, and tetrahedral or planar geometries (CN=4)~\cite{olofsson2000stabilization}. Additional lower-coordination configurations, such as triangular/T-shaped (CN=3)~\cite{olofsson1998first,bonhomme1992tetragonal} and linear (CN=2)~\cite{noreus1988na2pdh2}, have also been reported. This structural versatility, combined with very high hydrogen content, highlights the untapped potential of hydrido complexes for further exploration as superconducting materials. Nevertheless, only a handful (less than $\sim$10 \%) of the hundreds of known hydrido complexes are metallic~\cite{Yvon2006}, owing to the conventional electron counting rules usually obeyed by these compounds. Previous studies on the complex hydrides \ch{BaReH9}~\cite{muramatsu2015metallization} and \ch{Li5MoH11}~\cite{meng2019superconductivity} indicate metallization and superconductivity and with \Tc\ near \SI{7}{K} at very high pressures~\cite{muramatsu2015metallization}.

Recent computational predictions suggest ambient-pressure, high-\Tc\ superconductivity ranging from 65--\SI{170}{K} in the complex hydride \ch{Mg2IrH6}~\cite{sanna2024prediction,zheng2024prediction,dolui2024feasible}. \ch{Mg2IrH6} is metastable, but might be accessible through an insulating \ch{Mg2IrH7} intermediate phase predicted to be stable above \SI{15}{GPa}~\cite{dolui2024feasible}, or by hydrogenation of cubic \ch{Mg2IrH5}~\cite{hansen2024}. In parallel, \ch{Mg2PtH6} was also predicted to exhibit a high superconducting transition temperature in the range of 64--\SI{80}{K} and was suggested to be thermodynamically stable on the convex hull~\cite{sanna2024prediction}. These findings suggest significant potential for transition-metal hydride complexes beyond conventional hydrogen storage materials and demonstrate the possibility of high-\Tc\ hydride superconductivity within a new family of low-pressure materials. 

Using crystal structure prediction methods in combination with high-pressure and high-temperature (HPHT) experiments, we explored the phase behavior and compound formation in the Mg--Pt--H system. A new transition-metal hydride complex, \ch{Mg4Pt3H6}, was found to be the only thermodynamically stable ternary compound in this system and was synthesized using laser-heated diamond anvil cells (DACs) at pressures between $\sim$8--\SI{25}{GPa}. Synchrotron X-ray diffraction measurements were used to elucidate the structure of this body-centered cubic (BCC) hydride, which is comprised of Mg$^{2+}$ cations and linear [\ch{PtH2}]$^{2-}$ complexes. \ch{Mg4Pt3H6} is recoverable to ambient pressure with metallic conductivity and a superconducting transition with \Tc\ (50\%) = \SI{2.9}{K}.

\section{Methodology}

\textbf{Structure prediction and \textit{ab initio} calculations.} We employed crystal structure prediction (CSP) techniques \cite{reilly2016report} to determine low-energy phases within the Mg--Pt--H system. Structures within \SI{100}{meV/atom} of the ambient-pressure convex hull were considered from existing materials databases, in particular the Open Quantum Materials Database~\cite{saal2013materials}, Alexandria~\cite{schmidt2023machine}, and the Materials Project~\cite{jain2013commentary}. The accurate computational validation of the six known ground-state binary compounds (\ch{MgH2}, \ch{Mg3Pt}, \ch{Mg2Pt}, MgPt, \ch{MgPt3} and \ch{MgPt7}) ~\cite{nayeb1985mg,ellinger1955preparation,ferro1960research} demonstrates the robustness and reliability of our structure prediction methodologies. In order to streamline this process, we employed the standardized OPTIMADE API~\cite{evans2024developments}. These structures served as a starting point for training an ephemeral data-driven potential (EDDP)~\cite{pickard2022ephemeral} which was then combined with the \textit{ab initio} Random Structure Search (AIRSS) technique~\cite{pickard2011ab} to generate an additional 19,756 Mg--Pt--H structures, relaxed at \SI{10}{GPa} of external pressure to establish formation energies. Structures within \SI{100}{meV/atom} of this hull were then re-optimized at ambient pressure using DFT forces and stresses calculated with the \textsc{quantum espresso} software~\cite{giannozzi2009quantum}. Calculations employed the Perdew-Burke-Ernzerhof (PBE) exchange-correlation functional in the generalized gradient approximation (GGA)~\cite{perdew1996generalized}, v1.5 of the GBRV pseudopotentials~\cite{garrity2014pseudopotentials}, a Monkhorst-Pack~\cite{monkhorst1976special} $k$-grid with a spacing of at most \SI{0.04}{\r{A}}$^{-1}$, and a \SI{40}{Ry} plane-wave cutoff. Dynamic stability was examined using phonon spectrum calculations through the finite displacement method as implemented in the PHONOPY code~\cite{togo2008first}.

\textbf{Precursor preparation.} Initial hydride synthesis was attempted from a 2:1 mixture of Mg (Sigma Aldrich, 99.5$\%$) and Pt (Sigma Aldrich, 99.95$\%$) powders heated under 160 bars of high-purity hydrogen (Robert's Oxygen, 99.995\%) in a sealed autoclave-type reactor at 450 $^\circ$C for 8 days. Following heating, the reactor was cooled to ambient temperature and no ternary hydride phases were produced. Powder X-ray diffraction (XRD) analysis of the product revealed the formation of intermetallic \ch{Mg3Pt} with trace impurities of \ch{Mg2Pt} and \ch{Mg3Pt2} (Fig. S2). The product was found to be stable in air and was used as a precursor in subsequent high-pressure experiments. 

\textbf{Synthesis and characterization.} HPHT synthesis runs were conducted using symmetric, BX90-type, or non-magnetic miniature DACs with anvil culet diameters of 300--\SI{500}{\micro\meter}. Starting precursors were compressed into a pellet with a thickness of $\sim$5--\SI{10}{\micro\meter} and loaded into a sample chamber (diameter $\sim$150--\SI{250}{\micro\meter}) within a rhenium gasket that was pre-indented to a thickness of $\sim$\SI{40}{\micro\meter}. Starting precursor materials consisted of either the \ch{Mg3Pt} intermetallic phase described above, or well-mixed 2:1 Mg/Pt elemental powders (pressed as thin pellets). Samples containing elemental Mg were loaded inside a high-purity Ar glovebox to prevent possible oxidation. Subsequent to precursor loading, high-density fluid hydrogen (or ammonia borane for electrical transport samples) was loaded as a reagent and pressure-transmitting medium. 

Samples were compressed to various target pressures between 8.6--\SI{24.4}{GPa}, and one- or two-sided heating was performed using infrared laser heating systems (typically $\sim$\SI{100}{W}, \SI{1070}{nm} ytterbium fiber lasers). Pressures were estimated by the frequency shift of the diamond Raman edge~\cite{Akahama_2010} and through internal Pt or Au diffraction standards~\cite{anderson1989anharmonicity,holmes1989equation}. Temperature was estimated by fitting the thermal radiation spectra to the Planck radiation function from the heated sample position~\cite{Heinz1987}. In most cases, a chemical reaction was observed at low laser powers ($\sim$\SI{1600}{K}). XRD patterns for starting materials were measured using a Bruker D8 diffractometer (Cu-${K\alpha}$) equipped with a \SI{2}{\milli\meter} collimator and VANTEC-500 area detector. Focused monochromatic synchrotron XRD measurements were carried out in transmission geometry at beamline ID27~\cite{mezouar2024high} of the European Synchrotron Radiation Facility (runs 1--3, $\lambda$ = \SI{0.3738}{\AA}) using a Dectris Eiger 9M CdTe detector, at beamline I15 of the Diamond Light Source Facility (runs 4--5, $\lambda$ = \SI{0.4246}{\AA}) using a Eiger2 CdTe 9M detector, and at beamline 13-ID-D of GSECARS, Argonne National Laboratory (run 6, $\lambda$ = \SI{0.3344}{\AA}) using a Pilatus CdTe 9M detector. The sample--detector distance and other geometric parameters were calibrated using CeO$_2$ or \ch{LaB6} standards, and two-dimensional diffraction patterns were integrated using the DIOPTAS package~\cite{prescher2015dioptas}. For \textit{in situ} diffraction measurements, pressure determination was established using the equation of state (EOS) of Pt or Au (runs 1--4) and the Raman shift of diamond (run 5). Rietveld profile refinements were performed using the GSAS-II program~\cite{toby2013gsas}. The $P$--$V$ equation of state (EOS) for \ch{Mg4Pt3H6} was fitted using the third-order Birch--Murnaghan equation implemented in the EoSFit software package~\cite{gonzalez2016eosfit7}. 

\textbf{Electrical transport measurements.} Electrical transport measurements were performed using non-magnetic DACs with the Resistivity option of the Quantum Design Physical Property Measurement System (PPMS, Model 6000). Here, four Pt electrodes were cut from \SI{12}{\micro\meter} foil and placed within the sample chamber of an insulating gasket comprised of a c-BN--epoxy composite. A sample precursor pellet was loaded in contact with the electrodes and was surrounded by ammonia borane to release hydrogen upon heating. Four-probe resistance measurements with a maximum excitation current output of \SI{100}{\micro\A} were employed to eliminate the contact resistance of the electrodes, and temperature-dependent measurements between 1.9--\SI{300}{K} were performed before and after sample heating. Magnetic fields up to \SI{0.3}{T} were used to investigate the upper critical field. 

\section{Results and Discussion}

\subsection{Stable Mg--Pt--H Compounds} 
\begin{figure}[t!]
\begin{center}
\includegraphics[width=0.46\textwidth]{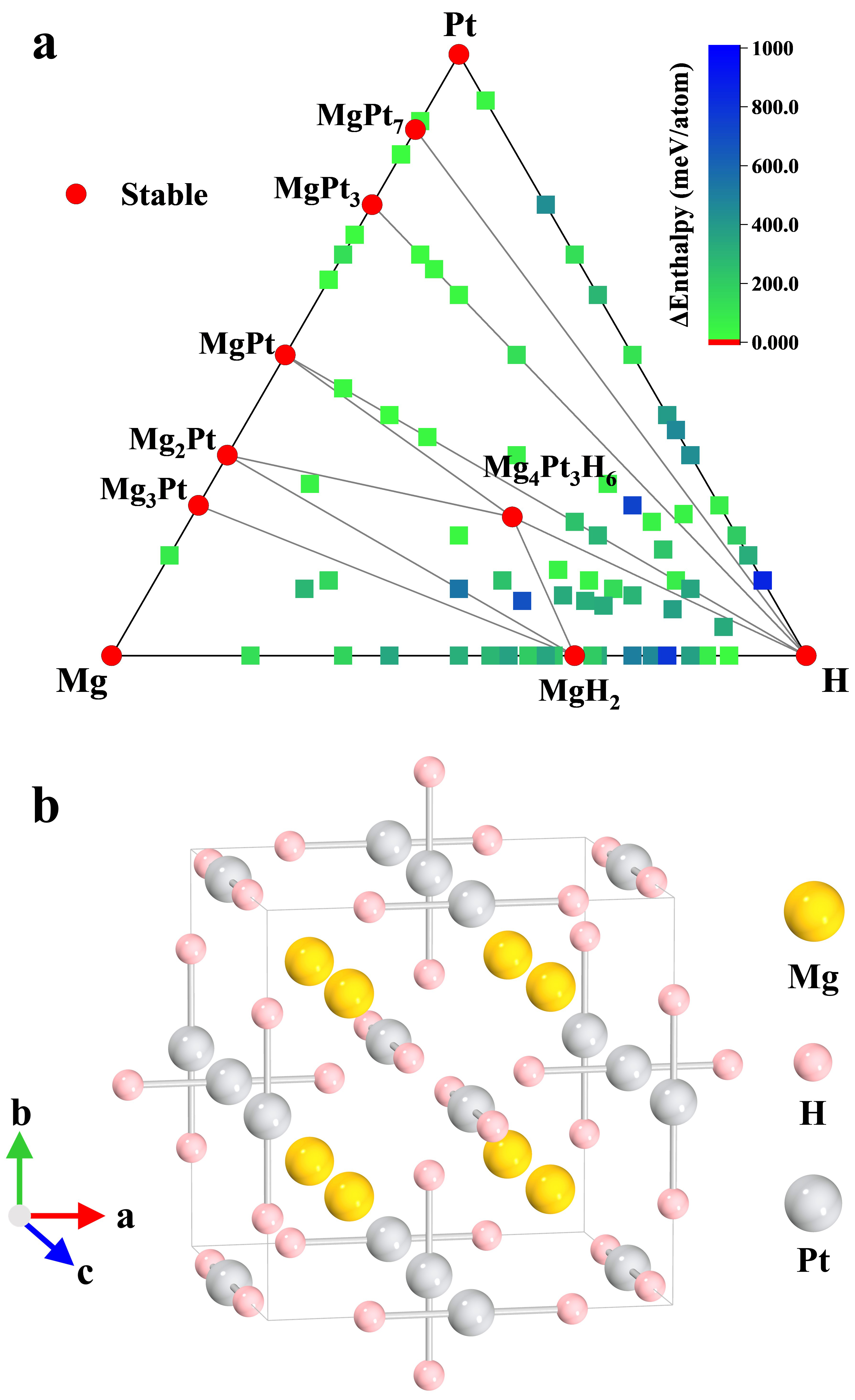}\\[5pt] 
\caption{\textit{\textbf{a}} Ternary phase diagram of the Mg--Pt--H system at ambient pressure based on DFT calculations. Various structures, represented by squares, are colored according to their distance from the convex hull, with red solid circles representing thermodynamically stable structures. \textit{Fm}$\Bar{3}$\textit{m} Pt, \textit{P}6$_3$/\textit{mmc} Mg, and \textit{Pa}$\Bar{3}$ \ch{H2} were used as elemental ground-state structures. \textit{\textbf{b}} The crystal structure of \textit{Im}$\Bar{3}$\textit{m} \ch{Mg4Pt3H6}. Mg, Pt and H atoms are shown as yellow, gray and pink spheres, respectively.}
\label{fig:Fig1}
\vspace{-0.8cm} 
\end{center}
\end{figure}

The ternary phase diagram of Mg--Pt--H at ambient pressure was constructed and visualized using the thermodynamic convex hull (Fig. \ref{fig:Fig1}a). This allowed the identification of thermodynamically stable Mg--Pt--H phases as those with negative formation enthalpy relative to all other phases (on-hull). A unique stoichiometric \ch{Mg4Pt3H6} compound was identified as the only ternary structure that remains on the convex hull from ambient pressure to at least 10 GPa. We note that \ch{Mg2PtH6}, previously estimated to be thermodynamically stable at low pressure~\cite{sanna2024prediction}, is predicted to lie $\sim$\SI{220}{meV/atom} above the convex hull at \SI{0}{GPa}. \ch{Mg4Pt3H6}  crystallizes in the body-centered cubic space group \textit{Im}$\Bar{3}$\textit{m} (No. 229; Z = 2) and contains Mg$^{2+}$ cations and linear [PtH$_2$]$^{2-}$ 14-electron complexes. The platinum atoms reside on the 6$b$ Wyckoff sites with 4/\textit{mm.m} symmetry and are coordinated by two linear hydrogen atoms occupying the 12$e$ Wyckoff positions with 4/\textit{m}.\textit{m} symmetry. Mg ions occupy the 8$c$ Wyckoff sites with .$\bar3 m$ symmetry (Fig. \ref{fig:Fig1}b). The calculated Pt--H bond length of \SI{1.657}{\AA} is similar to other known platinum hydrido complexes, with bond lengths typically ranging from 1.58 to \SI{1.66}{\AA}~\cite{Yvon2006}. The metal sublattice is similar to the Pt$_3$O$_4$ structure type, which contains Pt and O atoms on the 6$c$ and 8$b$ positions in space group \textit{Pm}$\Bar{3}$\textit{n} (No. 223), respectively. \ch{Mg4Pt3H6} is related to the complex hydride \ch{Ca8Rh6H24}, which contains the same metal sublattice and a three-dimensional network of corner-sharing octahedra with bridging and terminal H atoms (e.g., [Rh$_3$H$_{12}$]$_{n}^{8n-}$)~\cite{bronger1998calcium}, opposed to the linear [PtH$_2$]$^{2-}$ complex anions observed for \ch{Mg4Pt3H6}. Notably, \ch{Mg4Pt3H6} is predicted to be electron-imprecise with formal charge described as 4[Mg]$^{2+}\cdot$~3[PtH$_2$]$^{2-}$. While metallic complex transition-metal hydrides are known (see for example ref.~\cite{Yvon2006}), they are exceedingly uncommon, with only a handful of examples among the hundreds of known compounds.

Phonon dispersion calculations show that \textit{Im}$\Bar{3}$\textit{m} \ch{Mg4Pt3H6} is dynamically stable at ambient pressure, as evidenced by the absence of imaginary frequencies (Fig. S1) ~\cite{supp}. \textit{Im}$\Bar{3}$\textit{m} \ch{Mg4Pt3H6} exhibits the expected separation of hydrogen-dominant modes at high frequencies and heavier-element modes at lower frequencies. The predicted dynamic and thermodynamic stability of \ch{Mg4Pt3H6} at low pressure provides a solid foundation for experimental synthesis efforts.

\subsection{Synthesis of \textit{Im}$\Bar{3}$\textit{m} \ch{Mg4Pt3H6}} Following the systematic evaluation of stable phases in the Mg--Pt--H system, we conducted a series of experiments to validate computational predictions over a range of pressure--temperature conditions. Initial attempts to synthesize the target ternary compound by directly reacting well-mixed metal powders with hydrogen at 160 bar and 450 $^\circ$C in an autoclave-type reactor resulted only in the formation of Mg--Pt binaries, chiefly \ch{Mg3Pt} (Fig. S2)~\cite{supp}, with no hydrides produced. This observation suggests that \ch{Mg4Pt3H6} is not thermodyamically stable at ambient pressure under conditions of finite temperature, and indicates that entropic contributions are not negligible to the ambient-pressure convex hull. Nevertheless, \ch{Mg4Pt3H6} was predicted to exhibit thermodynamic stability to at least 10 GPa, which prompted additional synthesis experiments under elevated pressure conditions with the use of laser-heated DACs.

\begin{figure}[t!]
\begin{center}
\includegraphics[width=0.47\textwidth]{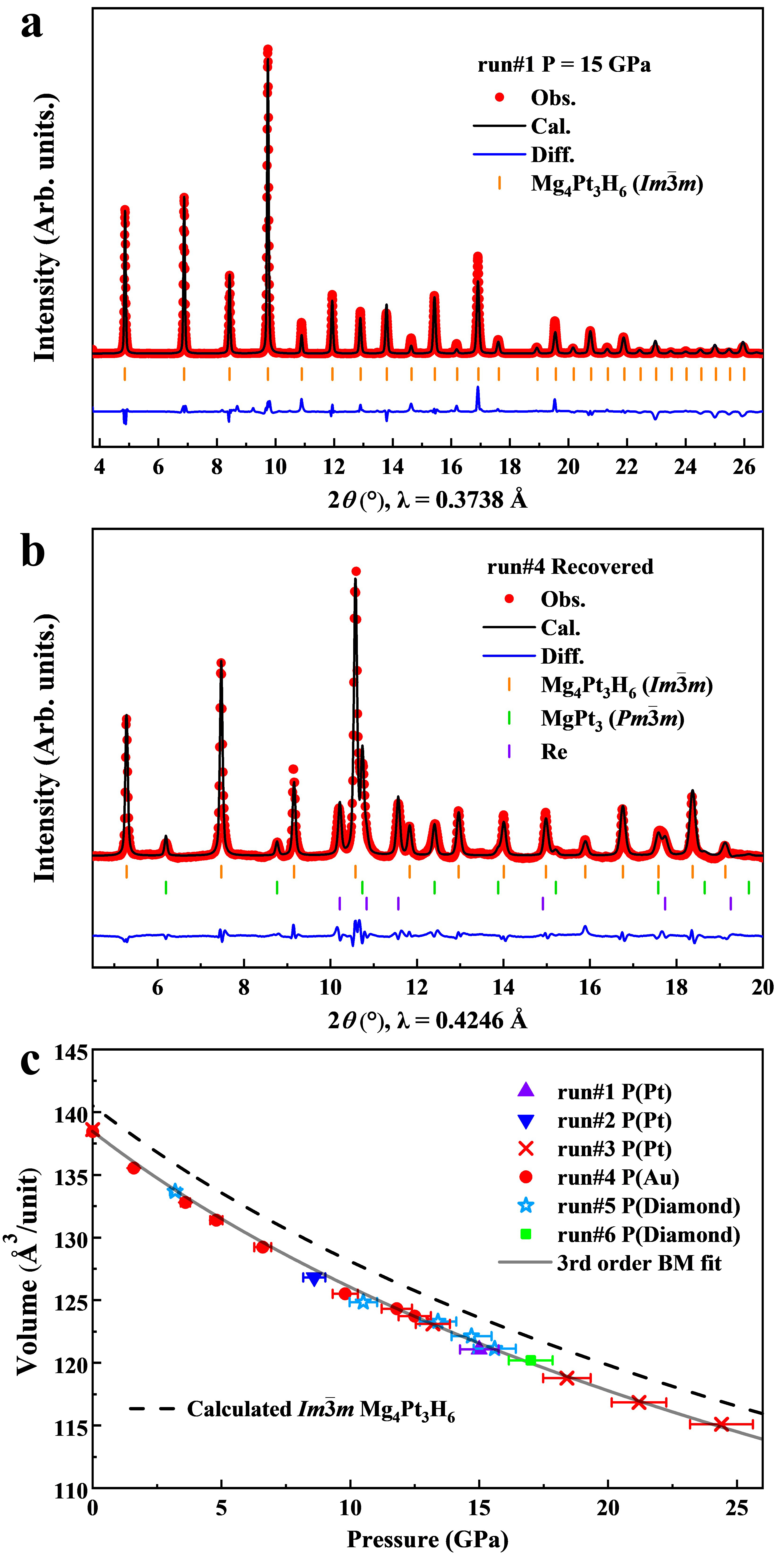}\\[5pt] 
\caption{\textbf{\textit{a}} Representative synchrotron powder XRD pattern (red points) for a sample synthesized at \SI{15}{GPa} (measured at room temperature) with Rietveld refinement (\textit{R}$_{wp}$ = 1.46\% and \textit{R}$_p$ = 2.83\%) of the predicted \textit{Im}$\Bar{3}$\textit{m} \ch{Mg4Pt3H6} structure (black line). The blue line shows the Rietveld residual and vertical ticks marks indicate allowed Bragg reflections. \textbf{\textit{b}} Rietveld refinement (\textit{R}$_{wp}$ = 2.48\% and \textit{R}$_p$ = 1.53\%) of the experimental XRD data at ambient pressure. Impurity peaks from \textit{Pm}$\bar3$\textit{m}-MgPt$_3$ and the Re gasket are represented by green and purple ticks, respectively. \textbf{\textit{c}} Pressure--Volume data for \textit{Im}$\Bar{3}$\textit{m} \ch{Mg4Pt3H6} obtained from all runs during decompression. The solid and dashed lines represent the experimental and DFT-PBE equations of state, respectively, as determined using the 3rd-order Birch--Murnaghan equation. Pressure markers for different experimental runs are indicated.}
\label{fig:Fig2}
\vspace{-0.8cm} 
\end{center}
\end{figure}

High-pressure synthesis experiments were conducted using intermetallic \ch{Mg3Pt} and homogeneous elemental metal powder precursors, reacted directly with molecular H$_2$ at pressures ranging between 8.6 to \SI{24.4}{GPa}. Upon laser heating samples to temperatures above $\sim$\SI{1600}{K}, new diffraction peaks were immediately observed using \textit{in situ} synchrotron powder XRD. In all six experimental runs, diffraction peaks were readily indexed to a BCC lattice with $a \approx$ 6.3 \AA, in good agreement with predictions for \ch{Mg4Pt3H6}. With additional laser annealing in runs 1, 5, and 6, phase-pure powder samples of this cubic phase were produced, as illustrated in Fig.\ref{fig:Fig2}a and Fig. S3. While the \textit{Im}$\Bar{3}$\textit{m} phase was the dominant product observed across all samples, we also observed \textit{Pm}$\Bar{3}$\textit{m}-\ch{MgPt3}~\cite{nayeb1985mg} and unreacted \ch{Mg3Pt} in some samples, likely reflecting local compositional and/or thermal gradients during heating. 

The synthesis of high-quality, phase-pure powder enabled the use of Rietveld refinement, as shown in Fig. \ref{fig:Fig2}a for a pattern obtained at \SI{15}{GPa}. The excellent agreement between observed powder diffraction intensities and the predicted \ch{Mg4Pt3H6} model confirms the metallic \textit{Im}$\Bar{3}$\textit{m} sublattice, and the refined lattice parameter of $a =$ \SI{6.233(4)}{\AA} at \SI{15}{GPa} is in close agreement with the DFT-predicted value of $a =$ \SI{6.276}{\AA}, further validating the structural prediction. While we are unable to determine hydrogen atomic positions from X-ray scattering, use of the calculated 12$e$ ($x$,0,0) position, in combination with the experimental lattice parameter, yields a Pt--H bond distance of 1.636 \AA~ at \SI{15}{GPa}. Experimentally and computationally determined crystallographic parameters for \ch{Mg4Pt3H6} are listed in Table S1~\cite{supp}.

To further examine the stability and compressibility of the synthesized \ch{Mg4Pt3H6} phase, we performed \textit{in situ} XRD measurements during decompression (Fig. S4) at room temperature across three independent experiments (runs 3--5). Notably, we also collected XRD patterns on two recovered samples at ambient pressure, as shown in Fig \ref{fig:Fig2}b and Fig. S4a. XRD patterns collected 12 hours after fully releasing the pressure confirmed the structural stability of \textit{Im}$\Bar{3}$\textit{m} \ch{Mg4Pt3H6} upon quenching to ambient pressure. These results confirm that the novel \ch{Mg4Pt3H6} compound is recoverable to ambient conditions, however, structural degradation of the \ch{Mg4Pt3H6} phase was observed after several months when the sample was stored in air. We note that many known hydrido complex compounds are sensitive to air and moisture.

The experimental $P$--$V$ data of \ch{Mg4Pt3H6} from all six experimental runs were fitted to a third-order Birch-Murnaghan equation of state (EOS) in the form
\begin{tiny}
$$P(V) = \frac{3}{2} B_0 \left[ \left(\frac{V_0}{V}\right)^{\frac{7}{3}} - \left(\frac{V_0}{V}\right)^{\frac{5}{3}} \right] 
\left\{ 1 + \frac{3}{4} \left(B'_0 - 4\right) \left[\left(\frac{V_0}{V}\right)^{\frac{2}{3}} - 1\right] \right\}.$$
\end{tiny}

The experimental EOS closely aligns with theoretical calculations, showing a volume deviation of 1.5--2\% over the entire pressure range, as shown in Fig. \ref{fig:Fig2}c, providing further confirmation of the synthesis of \ch{Mg4Pt3H6}. The slightly larger computational volumes obtained using the GGA method are consistent with trends observed in other materials, typically overestimating experimental values by 1--2\%~\cite{staroverov2004tests,wu2006more}. The experimental bulk modulus was determined to be B$_0$ = \SI{86.3(9)}{GPa} with a pressure derivative of B$_{0}^{’}$ = 4.39(8), in excellent agreement with the calculated values of B$_0$ = \SI{87.8}{GPa} and B$_{0}^{’}$ = 4.45. Given experimental limitations in determining quantitative hydrogen contents, we also explored the possibility for alternative Pt coordination schemes (e.g., [PtH$_4$]$^{2-}$) or interstitial hydrogen occupation and compared calculated volumes with experiment. No alternative compositions were found to exhibit better EOS agreement and no other compounds were found to be thermodynamically stable; thus, thermodynamically stable \ch{Mg4Pt3H6} is the only compound consistent with our experimental observations. 

\subsection{Electronic Structure} 
\begin{figure}[t!]
\begin{center}
\includegraphics[width=0.48\textwidth]{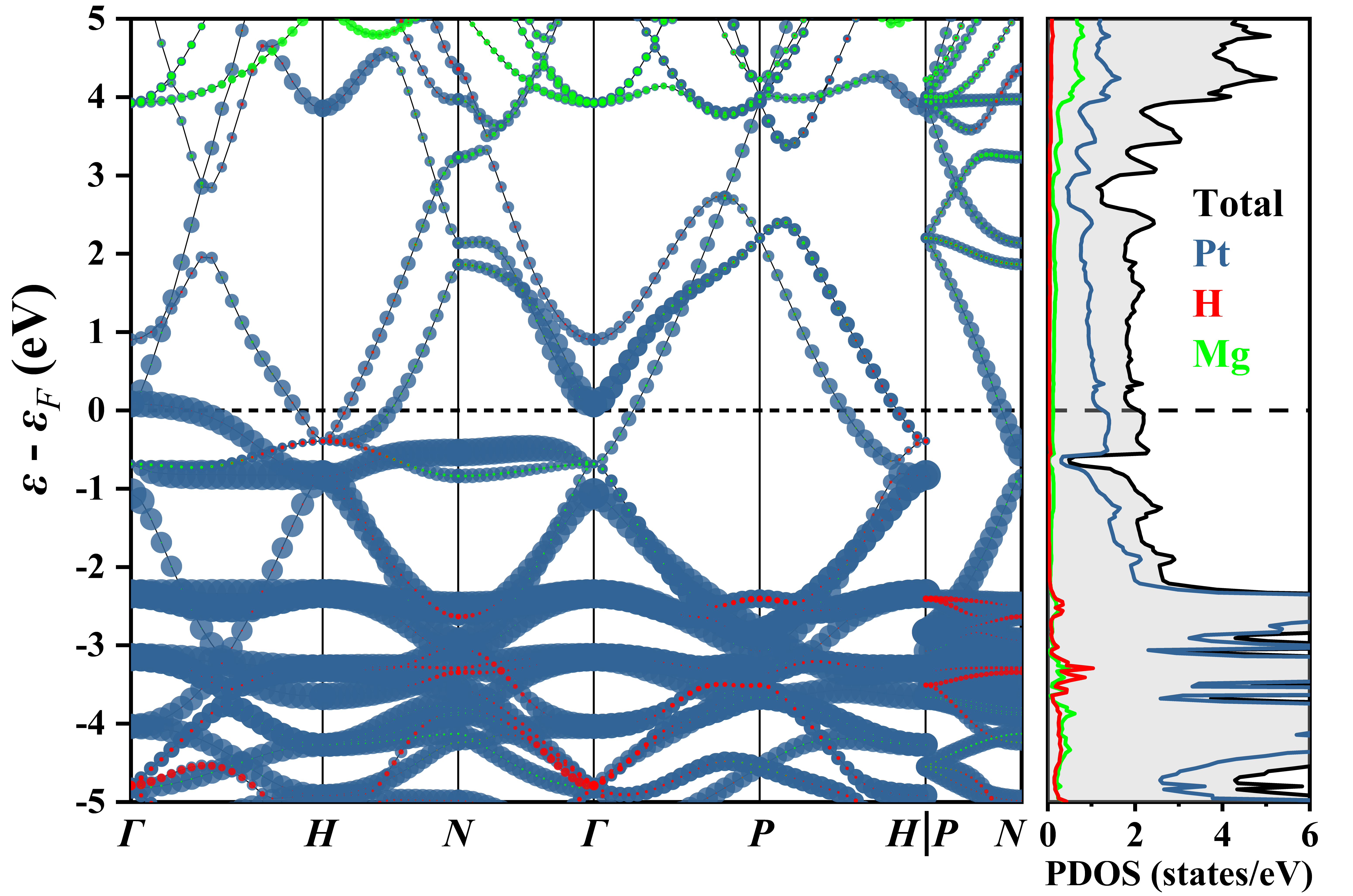}\\[5pt]  
\caption{Electronic band structure (left panel) and projected DOS (right panel) of \textit{Im}$\Bar{3}$\textit{m} \ch{Mg4Pt3H6} at ambient pressure. Point sizes indicate the respective projected weight and the dotted horizontal line represents the Fermi level.}
\label{fig:Fig3}
\vspace{-0.4cm} 
\end{center}
\end{figure}

Following the experimental validation of \ch{Mg4Pt3H6}, we performed additional theoretical and experimental studies of its electronic properties. The calculated electronic band structure and projected density of states (PDOS) of \ch{Mg4Pt3H6} at ambient pressure are presented in Fig. \ref{fig:Fig3}. As expected based on electron count, \ch{Mg4Pt3H6} is metallic with several bands crossing the Fermi level along high-symmetry directions. The PDOS shows that occupation at the Fermi level arises predominantly from $d$ and $p$ states of Pt, with small contributions from H and Mg atoms.  Electronic dispersion of platinum and hydrogen are also strongly correlated, reflecting hybridization in the [PtH$_2$]$^{2-}$ complex anions.  

Magnesium within \ch{Mg4Pt3H6} takes the valence state of +2, most common among other compounds. Bader charge analysis \cite{bader1991quantum} of the \textit{Im}$\Bar{3}$\textit{m} \ch{Mg4Pt3H6} structure at ambient pressure shows an average charge of 1.37 $e^-$ per H and 11.36 $e^-$ per Pt, indicating an average transfer of 2.1 $e^-$ to every PtH$_2$ complex from surrounding Mg. The formal charge description of 4[Mg]$^{2+}\cdot$~3[PtH$_2$]$^{2-}$ results in an imbalance of +2 per formula unit, giving rise to the metallic properties of the compound. This finding marks an unusual departure from the vast majority of previously known, platinum-based complex hydrides that are typically charge balanced~\cite{firman1998structure}. Among the previously reported hydrido complexes of Pt---for example, 2Sr$^{2+}\cdot$[PtH$_4$]$^{2^-}\cdot$2H$^{-}$~\cite{kadir1993a2h2}, 2Li$^{+}\cdot$[PtH$_2$]$^{2-}$~\cite{bronger1996li2pth2}, 2Na$^{+}\cdot$[PtH$_4$]$^{2-}$~\cite{bronger1984synthese},  3K$^{+}\cdot$[PtH$_4$]$^{2-}\cdot$H$^{-}$~\cite{bronger1988darstellung}, 2K$^{+}\cdot$Rb$^{+}\cdot$[PtH$_4$]$^{2-}\cdot$H$^{-}$~\cite{bronger1988darstellung}, 5Li$^+\cdot$[Pt$_2$H$_9$]$^{5-}$~\cite{bronger1995li5pt2h9}, and 2Na$^{+}\cdot$[PtH$_6$]$^{2-}$~\cite{bronger1995high}---we are unaware of a single known metallic compound. A small number of metallic complexes are known, \ch{Ca8Rh6H24} (4$n$Ca$^{2+}\cdot$[Rh$_3$H$_{12}$]$_{n}^{8n-}$, with idealized valence ``Rh$^{1.33}$'') being the most closely structurally related to \ch{Mg4Pt3H6}~\cite{bronger1998calcium}.

\begin{figure}[b!]
\begin{center}
\includegraphics[width=0.47\textwidth]{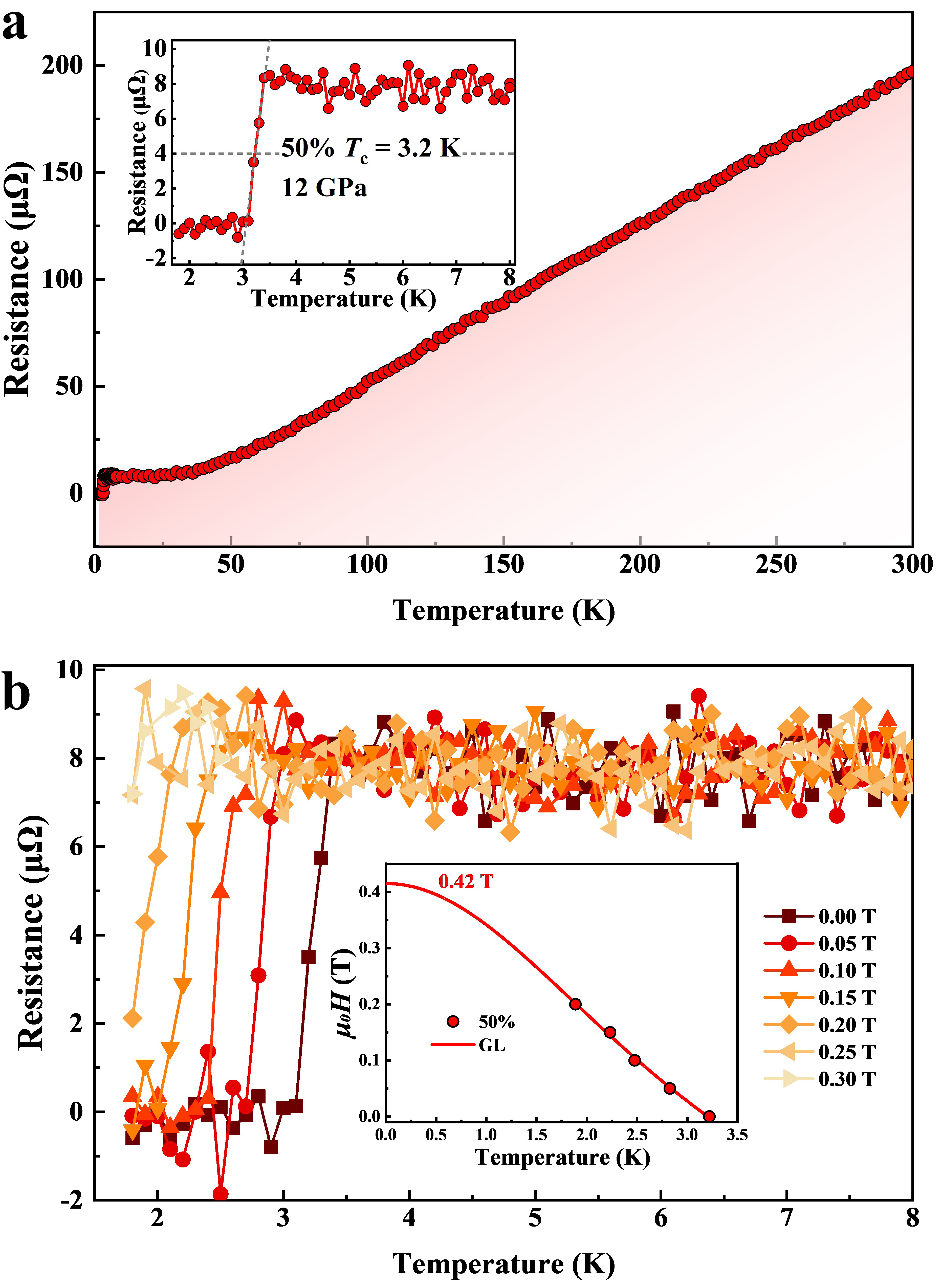}\\[5pt] 
\caption{\textbf{\textit{a}} Electrical transport measurements of the synthesized hydride under \SI{12}{GPa}. \textbf{\textit{b}} Temperature dependence of the electrical resistance under applied magnetic fields of $\mu_0 H = 0$--\SI{0.3}{T} at \SI{12}{GPa}. Inset: Upper critical field versus temperature following the criterion of 50\% of the resistance in the metallic state at \SI{12}{GPa}, fitted with the GL model~\cite{ginzburg2009theory}.}
\label{fig:Fig4}
\vspace{-0.4cm} 
\end{center}
\end{figure}

\subsection{Superconductivity} Following electronic structure calculations showing the metallic nature of \ch{Mg4Pt3H6}, we performed \textit{in situ} four-probe resistance measurements in non-magnetic DACs to understand electrical transport behavior. \ch{Mg3Pt} precursor pellets were laser heated near $\sim$\SI{15}{GPa} using ammonia borane as a hydrogen source, and the pressure decreased to \SI{12}{GPa} after heating. Before heating, the precursor phase exhibited metallic behavior with decreasing electrical resistance with decreasing temperature (Fig. S5). After heating, samples remain metallic, in agreement with the predicted metallic nature of \ch{Mg4Pt3H6}, however, a distinct drop to apparent zero resistance was observed with a midpoint temperature of \SI{3.2}{K} at \SI{12}{GPa} (Fig. \ref{fig:Fig4}a). This transition is suppressed under applied magnetic fields, indicative of a superconducting transition in the metallic complex hydride (Fig. \ref{fig:Fig4}b). By fitting the experimental midpoint transitions observed in the $R(T)$ data under applied magnetic fields up to \SI{0.2}{T} to a modified Ginzberg-Landau (GL) model~\cite{ginzburg2009theory} in the form $H_{c2}(T) = H_{c2}(0)(1-(\frac{T}{T_c})^2)$, where $H_{c2}(T)$ is the upper critical field and $T_c$ is the superconducting midpoint transition temperature, a value of \SI{0.42}{T} for $\mu_0H_{c2}(0)$ was obtained at \SI{12}{GPa}. The transition was maintained upon decompression to ambient pressure with a slight shift in the midpoint temperature to $T_c$ = \SI{2.9}{K} (Fig. S6), in agreement with the ambient-pressure recoverability of \ch{Mg4Pt3H6} based on X-ray diffraction measurements.

Electron--phonon interactions were calculated using density functional perturbation theory (DFPT) as implemented in \textsc{Quantum Espresso}~\cite{giannozzi2009quantum}, and $T_c$ was estimated using the Allen--Dynes-modified McMillan equation,~\cite{allen1975transition} assuming a Coulomb
pseudopotential value of $\mu^* = 0.1$. Calculations demonstrate that \ch{Mg4Pt3H6} is indeed superconducting, and we obtain an electron--phonon coupling parameter $\lambda=0.35$ and a logarithmic-average phonon frequency $\omega_{log} =$ \SI{493}{K}, resulting in an estimated $T_c =$ \SI{0.9}{K} at \SI{0}{GPa}. The small magnitude of $T_c$ arises due to a low density of electronic states at the Fermi level, the majority of which are localized to the heavier elements (see Fig.\ \ref{fig:Fig3}), resulting in weak electron--phonon coupling. The calculated superconducting transition temperature is lower than experimental values, but remains well within the expected accuracy for a weak-coupling superconductor. In this regime, the precise value of $T_c$ is very sensitive to the exponential dependence of the value of $\lambda$. Choice of the exchange--correlation functional, the value of $\mu^*$, the electronic smearing used to converge Brillouin zone integrals, or the treatment of anharmonicity can easily lead to variation of a few kelvin in $T_c$ (Fig. S7). Considering these factors, the calculated value is consistent with experiment and captures the correct physical behavior.
\\

\section{Conclusion}

In summary, we report the identification and synthesis of a stoichiometric complex transition-metal hydride, \textit{Im}$\bar{3}$\textit{m} \ch{Mg4Pt3H6}, via synergisitic crystal structure prediction and high-pressure experimental research. First-principles calculations, supported by thermodynamic and phonon stability analyses, reliably predicted the crystal structure and metallic electronic nature of the first known hydride in this ternary system. Experimental synthesis using laser-heated diamond anvil cells with X-ray diffraction confirmed the formation of the cubic structure and validated its compressibility. The hydride remains stable over a broad pressure range from \SI{25}{GPa} down to ambient pressure, as demonstrated by in situ pressure-quench measurements. Notably, electrical transport measurements revealed a superconducting transition with \Tc\ (50\%) = \SI{3.2}{K} at \SI{12}{GPa}, gradually decreasing to \SI{2.9}{K} at ambient pressure. These results indicate the possibility for a new class of low-pressure superconducting hydrides, and open new avenues for the design of complex hydrogen-rich materials with variable electronic properties.

\begin{acknowledgments}
We thank T. Bontke for assistance with XRD measurements and S. Liu for constructive discussions. This work was supported by the Enterprise Science Fund of Intellectual Ventures. We acknowledge the European Synchrotron Radiation Facility (ESRF) for provision of synchrotron radiation facilities at beamline ID27 (CH-6822)~\cite{2023_CH-6822_ESRF}. 
We acknowledge Diamond Light Source (DLS) for access and support of the beamline I15, proposal (CY35115--2).
Portions of this work were performed at GeoSoilEnviroCARS (The University of Chicago, Sector 13), Advanced Photon Source, Argonne National Laboratory. GeoSoilEnviroCARS is supported by the National Science Foundation – Earth Sciences via SEES: Synchrotron Earth and Environmental Science (EAR –2223273). This research used resources of the Advanced Photon Source, a U.S. Department of Energy (DOE) Office of Science User Facility operated for the DOE Office of Science by Argonne National Laboratory under Contract No. DE-AC02-06CH11357.
Use of the GSECARS Raman Lab System was supported by the NSF MRI Proposal (EAR-1531583). 
\end{acknowledgments}


\bibliography{ref}

\begin{thebibliography}{82}%
\makeatletter
\providecommand \@ifxundefined [1]{%
 \@ifx{#1\undefined}
}%
\providecommand \@ifnum [1]{%
 \ifnum #1\expandafter \@firstoftwo
 \else \expandafter \@secondoftwo
 \fi
}%
\providecommand \@ifx [1]{%
 \ifx #1\expandafter \@firstoftwo
 \else \expandafter \@secondoftwo
 \fi
}%
\providecommand \natexlab [1]{#1}%
\providecommand \enquote  [1]{``#1''}%
\providecommand \bibnamefont  [1]{#1}%
\providecommand \bibfnamefont [1]{#1}%
\providecommand \citenamefont [1]{#1}%
\providecommand \href@noop [0]{\@secondoftwo}%
\providecommand \href [0]{\begingroup \@sanitize@url \@href}%
\providecommand \@href[1]{\@@startlink{#1}\@@href}%
\providecommand \@@href[1]{\endgroup#1\@@endlink}%
\providecommand \@sanitize@url [0]{\catcode `\\12\catcode `\$12\catcode `\&12\catcode `\#12\catcode `\^12\catcode `\_12\catcode `\%12\relax}%
\providecommand \@@startlink[1]{}%
\providecommand \@@endlink[0]{}%
\providecommand \url  [0]{\begingroup\@sanitize@url \@url }%
\providecommand \@url [1]{\endgroup\@href {#1}{\urlprefix }}%
\providecommand \urlprefix  [0]{URL }%
\providecommand \Eprint [0]{\href }%
\providecommand \doibase [0]{https://doi.org/}%
\providecommand \selectlanguage [0]{\@gobble}%
\providecommand \bibinfo  [0]{\@secondoftwo}%
\providecommand \bibfield  [0]{\@secondoftwo}%
\providecommand \translation [1]{[#1]}%
\providecommand \BibitemOpen [0]{}%
\providecommand \bibitemStop [0]{}%
\providecommand \bibitemNoStop [0]{.\EOS\space}%
\providecommand \EOS [0]{\spacefactor3000\relax}%
\providecommand \BibitemShut  [1]{\csname bibitem#1\endcsname}%
\let\auto@bib@innerbib\@empty
\bibitem [{\citenamefont {Sun}\ \emph {et~al.}(2024)\citenamefont {Sun}, \citenamefont {Zhong}, \citenamefont {Liu},\ and\ \citenamefont {Ma}}]{sun2024clathrate}%
  \BibitemOpen
  \bibfield  {author} {\bibinfo {author} {\bibfnamefont {Y.}~\bibnamefont {Sun}}, \bibinfo {author} {\bibfnamefont {X.}~\bibnamefont {Zhong}}, \bibinfo {author} {\bibfnamefont {H.}~\bibnamefont {Liu}},\ and\ \bibinfo {author} {\bibfnamefont {Y.}~\bibnamefont {Ma}},\ }\bibfield  {title} {\bibinfo {title} {{Clathrate metal superhydrides under high-pressure conditions: enroute to room-temperature superconductivity}},\ }\href@noop {} {\bibfield  {journal} {\bibinfo  {journal} {Natl. Sci. Rev.}\ }\textbf {\bibinfo {volume} {11}},\ \bibinfo {pages} {nwad270} (\bibinfo {year} {2024})}\BibitemShut {NoStop}%
\bibitem [{\citenamefont {Gao}\ \emph {et~al.}(2021{\natexlab{a}})\citenamefont {Gao}, \citenamefont {Wang}, \citenamefont {Li}, \citenamefont {Zhang}, \citenamefont {Howie}, \citenamefont {Gregoryanz}, \citenamefont {Struzhkin}, \citenamefont {Wang},\ and\ \citenamefont {John}}]{gao2021superconducting}%
  \BibitemOpen
  \bibfield  {author} {\bibinfo {author} {\bibfnamefont {G.}~\bibnamefont {Gao}}, \bibinfo {author} {\bibfnamefont {L.}~\bibnamefont {Wang}}, \bibinfo {author} {\bibfnamefont {M.}~\bibnamefont {Li}}, \bibinfo {author} {\bibfnamefont {J.}~\bibnamefont {Zhang}}, \bibinfo {author} {\bibfnamefont {R.~T.}\ \bibnamefont {Howie}}, \bibinfo {author} {\bibfnamefont {E.}~\bibnamefont {Gregoryanz}}, \bibinfo {author} {\bibfnamefont {V.~V.}\ \bibnamefont {Struzhkin}}, \bibinfo {author} {\bibfnamefont {L.}~\bibnamefont {Wang}},\ and\ \bibinfo {author} {\bibfnamefont {S.~T.}\ \bibnamefont {John}},\ }\bibfield  {title} {\bibinfo {title} {{Superconducting binary hydrides: Theoretical predictions and experimental progresses}},\ }\href@noop {} {\bibfield  {journal} {\bibinfo  {journal} {Mater. Today Phys.}\ }\textbf {\bibinfo {volume} {21}},\ \bibinfo {pages} {100546} (\bibinfo {year} {2021}{\natexlab{a}})}\BibitemShut {NoStop}%
\bibitem [{\citenamefont {Drozdov}\ \emph {et~al.}(2015)\citenamefont {Drozdov}, \citenamefont {Eremets}, \citenamefont {Troyan}, \citenamefont {Ksenofontov},\ and\ \citenamefont {Shylin}}]{drozdov2015conventional}%
  \BibitemOpen
  \bibfield  {author} {\bibinfo {author} {\bibfnamefont {A.}~\bibnamefont {Drozdov}}, \bibinfo {author} {\bibfnamefont {M.}~\bibnamefont {Eremets}}, \bibinfo {author} {\bibfnamefont {I.}~\bibnamefont {Troyan}}, \bibinfo {author} {\bibfnamefont {V.}~\bibnamefont {Ksenofontov}},\ and\ \bibinfo {author} {\bibfnamefont {S.~I.}\ \bibnamefont {Shylin}},\ }\bibfield  {title} {\bibinfo {title} {{Conventional superconductivity at 203 kelvin at high pressures in the sulfur hydride system}},\ }\href@noop {} {\bibfield  {journal} {\bibinfo  {journal} {Nature}\ }\textbf {\bibinfo {volume} {525}},\ \bibinfo {pages} {73} (\bibinfo {year} {2015})}\BibitemShut {NoStop}%
\bibitem [{\citenamefont {Ma}\ \emph {et~al.}(2022)\citenamefont {Ma}, \citenamefont {Wang}, \citenamefont {Xie}, \citenamefont {Yang}, \citenamefont {Wang}, \citenamefont {Zhou}, \citenamefont {Liu}, \citenamefont {Yu}, \citenamefont {Zhao}, \citenamefont {Wang} \emph {et~al.}}]{ma2022high}%
  \BibitemOpen
  \bibfield  {author} {\bibinfo {author} {\bibfnamefont {L.}~\bibnamefont {Ma}}, \bibinfo {author} {\bibfnamefont {K.}~\bibnamefont {Wang}}, \bibinfo {author} {\bibfnamefont {Y.}~\bibnamefont {Xie}}, \bibinfo {author} {\bibfnamefont {X.}~\bibnamefont {Yang}}, \bibinfo {author} {\bibfnamefont {Y.}~\bibnamefont {Wang}}, \bibinfo {author} {\bibfnamefont {M.}~\bibnamefont {Zhou}}, \bibinfo {author} {\bibfnamefont {H.}~\bibnamefont {Liu}}, \bibinfo {author} {\bibfnamefont {X.}~\bibnamefont {Yu}}, \bibinfo {author} {\bibfnamefont {Y.}~\bibnamefont {Zhao}}, \bibinfo {author} {\bibfnamefont {H.}~\bibnamefont {Wang}}, \emph {et~al.},\ }\bibfield  {title} {\bibinfo {title} {{High-temperature superconducting phase in clathrate calcium hydride \ch{CaH6} up to \SI{215}{K} at a pressure of \SI{172}{GPa}}},\ }\href@noop {} {\bibfield  {journal} {\bibinfo  {journal} {Phys. Rev. Lett.}\ }\textbf {\bibinfo {volume} {128}},\ \bibinfo {pages} {167001} (\bibinfo {year} {2022})}\BibitemShut {NoStop}%
\bibitem [{\citenamefont {Wang}\ \emph {et~al.}(2012)\citenamefont {Wang}, \citenamefont {Tse}, \citenamefont {Tanaka}, \citenamefont {Iitaka},\ and\ \citenamefont {Ma}}]{wang2012superconductive}%
  \BibitemOpen
  \bibfield  {author} {\bibinfo {author} {\bibfnamefont {H.}~\bibnamefont {Wang}}, \bibinfo {author} {\bibfnamefont {J.~S.}\ \bibnamefont {Tse}}, \bibinfo {author} {\bibfnamefont {K.}~\bibnamefont {Tanaka}}, \bibinfo {author} {\bibfnamefont {T.}~\bibnamefont {Iitaka}},\ and\ \bibinfo {author} {\bibfnamefont {Y.}~\bibnamefont {Ma}},\ }\bibfield  {title} {\bibinfo {title} {{Superconductive sodalite-like clathrate calcium hydride at high pressures}},\ }\href@noop {} {\bibfield  {journal} {\bibinfo  {journal} {Proc. Natl. Acad. Sci. U.S.A.}\ }\textbf {\bibinfo {volume} {109}},\ \bibinfo {pages} {6463} (\bibinfo {year} {2012})}\BibitemShut {NoStop}%
\bibitem [{\citenamefont {Li}\ \emph {et~al.}(2022)\citenamefont {Li}, \citenamefont {He}, \citenamefont {Zhang}, \citenamefont {Wang}, \citenamefont {Zhang}, \citenamefont {Jia}, \citenamefont {Feng}, \citenamefont {Lu}, \citenamefont {Zhao}, \citenamefont {Zhang} \emph {et~al.}}]{li2022superconductivity}%
  \BibitemOpen
  \bibfield  {author} {\bibinfo {author} {\bibfnamefont {Z.}~\bibnamefont {Li}}, \bibinfo {author} {\bibfnamefont {X.}~\bibnamefont {He}}, \bibinfo {author} {\bibfnamefont {C.}~\bibnamefont {Zhang}}, \bibinfo {author} {\bibfnamefont {X.}~\bibnamefont {Wang}}, \bibinfo {author} {\bibfnamefont {S.}~\bibnamefont {Zhang}}, \bibinfo {author} {\bibfnamefont {Y.}~\bibnamefont {Jia}}, \bibinfo {author} {\bibfnamefont {S.}~\bibnamefont {Feng}}, \bibinfo {author} {\bibfnamefont {K.}~\bibnamefont {Lu}}, \bibinfo {author} {\bibfnamefont {J.}~\bibnamefont {Zhao}}, \bibinfo {author} {\bibfnamefont {J.}~\bibnamefont {Zhang}}, \emph {et~al.},\ }\bibfield  {title} {\bibinfo {title} {{Superconductivity above \SI{200}{K} discovered in superhydrides of calcium}},\ }\href@noop {} {\bibfield  {journal} {\bibinfo  {journal} {Nat. Commun.}\ }\textbf {\bibinfo {volume} {13}},\ \bibinfo {pages} {2863} (\bibinfo {year} {2022})}\BibitemShut {NoStop}%
\bibitem [{\citenamefont {Kong}\ \emph {et~al.}(2021)\citenamefont {Kong}, \citenamefont {Minkov}, \citenamefont {Kuzovnikov}, \citenamefont {Drozdov}, \citenamefont {Besedin}, \citenamefont {Mozaffari}, \citenamefont {Balicas}, \citenamefont {Balakirev}, \citenamefont {Prakapenka}, \citenamefont {Chariton} \emph {et~al.}}]{kong2021superconductivity}%
  \BibitemOpen
  \bibfield  {author} {\bibinfo {author} {\bibfnamefont {P.}~\bibnamefont {Kong}}, \bibinfo {author} {\bibfnamefont {V.~S.}\ \bibnamefont {Minkov}}, \bibinfo {author} {\bibfnamefont {M.~A.}\ \bibnamefont {Kuzovnikov}}, \bibinfo {author} {\bibfnamefont {A.~P.}\ \bibnamefont {Drozdov}}, \bibinfo {author} {\bibfnamefont {S.~P.}\ \bibnamefont {Besedin}}, \bibinfo {author} {\bibfnamefont {S.}~\bibnamefont {Mozaffari}}, \bibinfo {author} {\bibfnamefont {L.}~\bibnamefont {Balicas}}, \bibinfo {author} {\bibfnamefont {F.~F.}\ \bibnamefont {Balakirev}}, \bibinfo {author} {\bibfnamefont {V.~B.}\ \bibnamefont {Prakapenka}}, \bibinfo {author} {\bibfnamefont {S.}~\bibnamefont {Chariton}}, \emph {et~al.},\ }\bibfield  {title} {\bibinfo {title} {{Superconductivity up to \SI{243}{K} in the yttrium-hydrogen system under high pressure}},\ }\href@noop {} {\bibfield  {journal} {\bibinfo  {journal} {Nat. Commun.}\ }\textbf {\bibinfo {volume} {12}},\ \bibinfo {pages} {5075} (\bibinfo {year} {2021})}\BibitemShut {NoStop}%
\bibitem [{\citenamefont {Troyan}\ \emph {et~al.}(2021)\citenamefont {Troyan}, \citenamefont {Semenok}, \citenamefont {Kvashnin}, \citenamefont {Sadakov}, \citenamefont {Sobolevskiy}, \citenamefont {Pudalov}, \citenamefont {Ivanova}, \citenamefont {Prakapenka}, \citenamefont {Greenberg}, \citenamefont {Gavriliuk} \emph {et~al.}}]{troyan2021anomalous}%
  \BibitemOpen
  \bibfield  {author} {\bibinfo {author} {\bibfnamefont {I.~A.}\ \bibnamefont {Troyan}}, \bibinfo {author} {\bibfnamefont {D.~V.}\ \bibnamefont {Semenok}}, \bibinfo {author} {\bibfnamefont {A.~G.}\ \bibnamefont {Kvashnin}}, \bibinfo {author} {\bibfnamefont {A.~V.}\ \bibnamefont {Sadakov}}, \bibinfo {author} {\bibfnamefont {O.~A.}\ \bibnamefont {Sobolevskiy}}, \bibinfo {author} {\bibfnamefont {V.~M.}\ \bibnamefont {Pudalov}}, \bibinfo {author} {\bibfnamefont {A.~G.}\ \bibnamefont {Ivanova}}, \bibinfo {author} {\bibfnamefont {V.~B.}\ \bibnamefont {Prakapenka}}, \bibinfo {author} {\bibfnamefont {E.}~\bibnamefont {Greenberg}}, \bibinfo {author} {\bibfnamefont {A.~G.}\ \bibnamefont {Gavriliuk}}, \emph {et~al.},\ }\bibfield  {title} {\bibinfo {title} {{Anomalous high-temperature superconductivity in \ch{YH6}}},\ }\href@noop {} {\bibfield  {journal} {\bibinfo  {journal} {Adv. Mater.}\ }\textbf {\bibinfo {volume} {33}},\ \bibinfo {pages} {2006832} (\bibinfo {year} {2021})}\BibitemShut {NoStop}%
\bibitem [{\citenamefont {Peng}\ \emph {et~al.}(2017)\citenamefont {Peng}, \citenamefont {Sun}, \citenamefont {Pickard}, \citenamefont {Needs}, \citenamefont {Wu},\ and\ \citenamefont {Ma}}]{peng2017hydrogen}%
  \BibitemOpen
  \bibfield  {author} {\bibinfo {author} {\bibfnamefont {F.}~\bibnamefont {Peng}}, \bibinfo {author} {\bibfnamefont {Y.}~\bibnamefont {Sun}}, \bibinfo {author} {\bibfnamefont {C.~J.}\ \bibnamefont {Pickard}}, \bibinfo {author} {\bibfnamefont {R.~J.}\ \bibnamefont {Needs}}, \bibinfo {author} {\bibfnamefont {Q.}~\bibnamefont {Wu}},\ and\ \bibinfo {author} {\bibfnamefont {Y.}~\bibnamefont {Ma}},\ }\bibfield  {title} {\bibinfo {title} {{Hydrogen clathrate structures in rare earth hydrides at high pressures: possible route to room-temperature superconductivity}},\ }\href@noop {} {\bibfield  {journal} {\bibinfo  {journal} {Phys. Rev. Lett.}\ }\textbf {\bibinfo {volume} {119}},\ \bibinfo {pages} {107001} (\bibinfo {year} {2017})}\BibitemShut {NoStop}%
\bibitem [{\citenamefont {Snider}\ \emph {et~al.}(2021)\citenamefont {Snider}, \citenamefont {Dasenbrock-Gammon}, \citenamefont {McBride}, \citenamefont {Wang}, \citenamefont {Meyers}, \citenamefont {Lawler}, \citenamefont {Zurek}, \citenamefont {Salamat},\ and\ \citenamefont {Dias}}]{snider2021synthesis}%
  \BibitemOpen
  \bibfield  {author} {\bibinfo {author} {\bibfnamefont {E.}~\bibnamefont {Snider}}, \bibinfo {author} {\bibfnamefont {N.}~\bibnamefont {Dasenbrock-Gammon}}, \bibinfo {author} {\bibfnamefont {R.}~\bibnamefont {McBride}}, \bibinfo {author} {\bibfnamefont {X.}~\bibnamefont {Wang}}, \bibinfo {author} {\bibfnamefont {N.}~\bibnamefont {Meyers}}, \bibinfo {author} {\bibfnamefont {K.~V.}\ \bibnamefont {Lawler}}, \bibinfo {author} {\bibfnamefont {E.}~\bibnamefont {Zurek}}, \bibinfo {author} {\bibfnamefont {A.}~\bibnamefont {Salamat}},\ and\ \bibinfo {author} {\bibfnamefont {R.~P.}\ \bibnamefont {Dias}},\ }\bibfield  {title} {\bibinfo {title} {{Synthesis of yttrium superhydride superconductor with a transition temperature up to \SI{262}{K} by catalytic hydrogenation at high pressures}},\ }\href@noop {} {\bibfield  {journal} {\bibinfo  {journal} {Phys. Rev. Lett.}\ }\textbf {\bibinfo {volume} {126}},\ \bibinfo {pages} {117003} (\bibinfo {year} {2021})}\BibitemShut {NoStop}%
\bibitem [{\citenamefont {Liu}\ \emph {et~al.}(2017)\citenamefont {Liu}, \citenamefont {Naumov}, \citenamefont {Hoffmann}, \citenamefont {Ashcroft},\ and\ \citenamefont {Hemley}}]{liu2017potential}%
  \BibitemOpen
  \bibfield  {author} {\bibinfo {author} {\bibfnamefont {H.}~\bibnamefont {Liu}}, \bibinfo {author} {\bibfnamefont {I.~I.}\ \bibnamefont {Naumov}}, \bibinfo {author} {\bibfnamefont {R.}~\bibnamefont {Hoffmann}}, \bibinfo {author} {\bibfnamefont {N.}~\bibnamefont {Ashcroft}},\ and\ \bibinfo {author} {\bibfnamefont {R.~J.}\ \bibnamefont {Hemley}},\ }\bibfield  {title} {\bibinfo {title} {{Potential high-\Tc\ superconducting lanthanum and yttrium hydrides at high pressure}},\ }\href@noop {} {\bibfield  {journal} {\bibinfo  {journal} {Proc. Natl. Acad. Sci. U.S.A.}\ }\textbf {\bibinfo {volume} {114}},\ \bibinfo {pages} {6990} (\bibinfo {year} {2017})}\BibitemShut {NoStop}%
\bibitem [{\citenamefont {Drozdov}\ \emph {et~al.}(2019)\citenamefont {Drozdov}, \citenamefont {Kong}, \citenamefont {Minkov}, \citenamefont {Besedin}, \citenamefont {Kuzovnikov}, \citenamefont {Mozaffari}, \citenamefont {Balicas}, \citenamefont {Balakirev}, \citenamefont {Graf}, \citenamefont {Prakapenka} \emph {et~al.}}]{drozdov2019superconductivity}%
  \BibitemOpen
  \bibfield  {author} {\bibinfo {author} {\bibfnamefont {A.}~\bibnamefont {Drozdov}}, \bibinfo {author} {\bibfnamefont {P.}~\bibnamefont {Kong}}, \bibinfo {author} {\bibfnamefont {V.}~\bibnamefont {Minkov}}, \bibinfo {author} {\bibfnamefont {S.}~\bibnamefont {Besedin}}, \bibinfo {author} {\bibfnamefont {M.}~\bibnamefont {Kuzovnikov}}, \bibinfo {author} {\bibfnamefont {S.}~\bibnamefont {Mozaffari}}, \bibinfo {author} {\bibfnamefont {L.}~\bibnamefont {Balicas}}, \bibinfo {author} {\bibfnamefont {F.}~\bibnamefont {Balakirev}}, \bibinfo {author} {\bibfnamefont {D.}~\bibnamefont {Graf}}, \bibinfo {author} {\bibfnamefont {V.}~\bibnamefont {Prakapenka}}, \emph {et~al.},\ }\bibfield  {title} {\bibinfo {title} {{Superconductivity at \SI{250}{K} in lanthanum hydride under high pressures}},\ }\href@noop {} {\bibfield  {journal} {\bibinfo  {journal} {Nature}\ }\textbf {\bibinfo {volume} {569}},\ \bibinfo {pages} {528} (\bibinfo {year} {2019})}\BibitemShut {NoStop}%
\bibitem [{\citenamefont {Somayazulu}\ \emph {et~al.}(2019)\citenamefont {Somayazulu}, \citenamefont {Ahart}, \citenamefont {Mishra}, \citenamefont {Geballe}, \citenamefont {Baldini}, \citenamefont {Meng}, \citenamefont {Struzhkin},\ and\ \citenamefont {Hemley}}]{somayazulu2019evidence}%
  \BibitemOpen
  \bibfield  {author} {\bibinfo {author} {\bibfnamefont {M.}~\bibnamefont {Somayazulu}}, \bibinfo {author} {\bibfnamefont {M.}~\bibnamefont {Ahart}}, \bibinfo {author} {\bibfnamefont {A.~K.}\ \bibnamefont {Mishra}}, \bibinfo {author} {\bibfnamefont {Z.~M.}\ \bibnamefont {Geballe}}, \bibinfo {author} {\bibfnamefont {M.}~\bibnamefont {Baldini}}, \bibinfo {author} {\bibfnamefont {Y.}~\bibnamefont {Meng}}, \bibinfo {author} {\bibfnamefont {V.~V.}\ \bibnamefont {Struzhkin}},\ and\ \bibinfo {author} {\bibfnamefont {R.~J.}\ \bibnamefont {Hemley}},\ }\bibfield  {title} {\bibinfo {title} {{Evidence for superconductivity above \SI{260}{K} in lanthanum superhydride at megabar pressures}},\ }\href@noop {} {\bibfield  {journal} {\bibinfo  {journal} {Phys. Rev. Lett.}\ }\textbf {\bibinfo {volume} {122}},\ \bibinfo {pages} {027001} (\bibinfo {year} {2019})}\BibitemShut {NoStop}%
\bibitem [{\citenamefont {Zhang}\ \emph {et~al.}(2022{\natexlab{a}})\citenamefont {Zhang}, \citenamefont {Zhao},\ and\ \citenamefont {Yang}}]{zhang2022superconducting}%
  \BibitemOpen
  \bibfield  {author} {\bibinfo {author} {\bibfnamefont {X.}~\bibnamefont {Zhang}}, \bibinfo {author} {\bibfnamefont {Y.}~\bibnamefont {Zhao}},\ and\ \bibinfo {author} {\bibfnamefont {G.}~\bibnamefont {Yang}},\ }\bibfield  {title} {\bibinfo {title} {{Superconducting ternary hydrides under high pressure}},\ }\href@noop {} {\bibfield  {journal} {\bibinfo  {journal} {WIREs Comput. Mol. Sci.}\ }\textbf {\bibinfo {volume} {12}},\ \bibinfo {pages} {e1582} (\bibinfo {year} {2022}{\natexlab{a}})}\BibitemShut {NoStop}%
\bibitem [{\citenamefont {Semenok}\ \emph {et~al.}(2021)\citenamefont {Semenok}, \citenamefont {Troyan}, \citenamefont {Ivanova}, \citenamefont {Kvashnin}, \citenamefont {Kruglov}, \citenamefont {Hanfland}, \citenamefont {Sadakov}, \citenamefont {Sobolevskiy}, \citenamefont {Pervakov}, \citenamefont {Lyubutin} \emph {et~al.}}]{semenok2021superconductivity}%
  \BibitemOpen
  \bibfield  {author} {\bibinfo {author} {\bibfnamefont {D.~V.}\ \bibnamefont {Semenok}}, \bibinfo {author} {\bibfnamefont {I.~A.}\ \bibnamefont {Troyan}}, \bibinfo {author} {\bibfnamefont {A.~G.}\ \bibnamefont {Ivanova}}, \bibinfo {author} {\bibfnamefont {A.~G.}\ \bibnamefont {Kvashnin}}, \bibinfo {author} {\bibfnamefont {I.~A.}\ \bibnamefont {Kruglov}}, \bibinfo {author} {\bibfnamefont {M.}~\bibnamefont {Hanfland}}, \bibinfo {author} {\bibfnamefont {A.~V.}\ \bibnamefont {Sadakov}}, \bibinfo {author} {\bibfnamefont {O.~A.}\ \bibnamefont {Sobolevskiy}}, \bibinfo {author} {\bibfnamefont {K.~S.}\ \bibnamefont {Pervakov}}, \bibinfo {author} {\bibfnamefont {I.~S.}\ \bibnamefont {Lyubutin}}, \emph {et~al.},\ }\bibfield  {title} {\bibinfo {title} {{Superconductivity at \SI{253}{K} in lanthanum--yttrium ternary hydrides}},\ }\href@noop {} {\bibfield  {journal} {\bibinfo  {journal} {Mater. Today}\ }\textbf {\bibinfo {volume} {48}},\ \bibinfo {pages} {18} (\bibinfo {year} {2021})}\BibitemShut {NoStop}%
\bibitem [{\citenamefont {Chen}\ \emph {et~al.}(2024{\natexlab{a}})\citenamefont {Chen}, \citenamefont {Luo}, \citenamefont {Cao}, \citenamefont {Dalladay-Simpson}, \citenamefont {Huang}, \citenamefont {Peng}, \citenamefont {Zhang}, \citenamefont {Gorelli}, \citenamefont {Zhong}, \citenamefont {Lin} \emph {et~al.}}]{chen2024synthesis}%
  \BibitemOpen
  \bibfield  {author} {\bibinfo {author} {\bibfnamefont {L.-C.}\ \bibnamefont {Chen}}, \bibinfo {author} {\bibfnamefont {T.}~\bibnamefont {Luo}}, \bibinfo {author} {\bibfnamefont {Z.-Y.}\ \bibnamefont {Cao}}, \bibinfo {author} {\bibfnamefont {P.}~\bibnamefont {Dalladay-Simpson}}, \bibinfo {author} {\bibfnamefont {G.}~\bibnamefont {Huang}}, \bibinfo {author} {\bibfnamefont {D.}~\bibnamefont {Peng}}, \bibinfo {author} {\bibfnamefont {L.-L.}\ \bibnamefont {Zhang}}, \bibinfo {author} {\bibfnamefont {F.~A.}\ \bibnamefont {Gorelli}}, \bibinfo {author} {\bibfnamefont {G.-H.}\ \bibnamefont {Zhong}}, \bibinfo {author} {\bibfnamefont {H.-Q.}\ \bibnamefont {Lin}}, \emph {et~al.},\ }\bibfield  {title} {\bibinfo {title} {{Synthesis and superconductivity in yttrium-cerium hydrides at high pressures}},\ }\href@noop {} {\bibfield  {journal} {\bibinfo  {journal} {Nat. Commun.}\ }\textbf {\bibinfo {volume} {15}},\ \bibinfo {pages} {1809} (\bibinfo {year} {2024}{\natexlab{a}})}\BibitemShut {NoStop}%
\bibitem [{\citenamefont {Chen}\ \emph {et~al.}(2024{\natexlab{b}})\citenamefont {Chen}, \citenamefont {Qian}, \citenamefont {Huang}, \citenamefont {Chen}, \citenamefont {Guo}, \citenamefont {Zhang}, \citenamefont {Zhang}, \citenamefont {Yuan},\ and\ \citenamefont {Cui}}]{chen2024high}%
  \BibitemOpen
  \bibfield  {author} {\bibinfo {author} {\bibfnamefont {S.}~\bibnamefont {Chen}}, \bibinfo {author} {\bibfnamefont {Y.}~\bibnamefont {Qian}}, \bibinfo {author} {\bibfnamefont {X.}~\bibnamefont {Huang}}, \bibinfo {author} {\bibfnamefont {W.}~\bibnamefont {Chen}}, \bibinfo {author} {\bibfnamefont {J.}~\bibnamefont {Guo}}, \bibinfo {author} {\bibfnamefont {K.}~\bibnamefont {Zhang}}, \bibinfo {author} {\bibfnamefont {J.}~\bibnamefont {Zhang}}, \bibinfo {author} {\bibfnamefont {H.}~\bibnamefont {Yuan}},\ and\ \bibinfo {author} {\bibfnamefont {T.}~\bibnamefont {Cui}},\ }\bibfield  {title} {\bibinfo {title} {{High-temperature superconductivity up to \SI{233}{K} in the Al stabilized metastable hexagonal lanthanum superhydride}},\ }\href@noop {} {\bibfield  {journal} {\bibinfo  {journal} {Natl. Sci. Rev.}\ }\textbf {\bibinfo {volume} {11}},\ \bibinfo {pages} {nwad107} (\bibinfo {year} {2024}{\natexlab{b}})}\BibitemShut {NoStop}%
\bibitem [{\citenamefont {Bi}\ \emph {et~al.}(2022)\citenamefont {Bi}, \citenamefont {Nakamoto}, \citenamefont {Zhang}, \citenamefont {Shimizu}, \citenamefont {Zou}, \citenamefont {Liu}, \citenamefont {Zhou}, \citenamefont {Liu}, \citenamefont {Wang},\ and\ \citenamefont {Ma}}]{bi2022giant}%
  \BibitemOpen
  \bibfield  {author} {\bibinfo {author} {\bibfnamefont {J.}~\bibnamefont {Bi}}, \bibinfo {author} {\bibfnamefont {Y.}~\bibnamefont {Nakamoto}}, \bibinfo {author} {\bibfnamefont {P.}~\bibnamefont {Zhang}}, \bibinfo {author} {\bibfnamefont {K.}~\bibnamefont {Shimizu}}, \bibinfo {author} {\bibfnamefont {B.}~\bibnamefont {Zou}}, \bibinfo {author} {\bibfnamefont {H.}~\bibnamefont {Liu}}, \bibinfo {author} {\bibfnamefont {M.}~\bibnamefont {Zhou}}, \bibinfo {author} {\bibfnamefont {G.}~\bibnamefont {Liu}}, \bibinfo {author} {\bibfnamefont {H.}~\bibnamefont {Wang}},\ and\ \bibinfo {author} {\bibfnamefont {Y.}~\bibnamefont {Ma}},\ }\bibfield  {title} {\bibinfo {title} {{Giant enhancement of superconducting critical temperature in substitutional alloy \ch{(La,Ce)H9}}},\ }\href@noop {} {\bibfield  {journal} {\bibinfo  {journal} {Nat. Commun.}\ }\textbf {\bibinfo {volume} {13}},\ \bibinfo {pages} {5952} (\bibinfo {year} {2022})}\BibitemShut {NoStop}%
\bibitem [{\citenamefont {Zhang}\ \emph {et~al.}(2025)\citenamefont {Zhang}, \citenamefont {Yu}, \citenamefont {Zhang}, \citenamefont {Guo}, \citenamefont {Wang}, \citenamefont {Jiang}, \citenamefont {Huang},\ and\ \citenamefont {Cui}}]{zhang2025synthesis}%
  \BibitemOpen
  \bibfield  {author} {\bibinfo {author} {\bibfnamefont {K.}~\bibnamefont {Zhang}}, \bibinfo {author} {\bibfnamefont {J.}~\bibnamefont {Yu}}, \bibinfo {author} {\bibfnamefont {Y.}~\bibnamefont {Zhang}}, \bibinfo {author} {\bibfnamefont {J.}~\bibnamefont {Guo}}, \bibinfo {author} {\bibfnamefont {Y.}~\bibnamefont {Wang}}, \bibinfo {author} {\bibfnamefont {C.}~\bibnamefont {Jiang}}, \bibinfo {author} {\bibfnamefont {X.}~\bibnamefont {Huang}},\ and\ \bibinfo {author} {\bibfnamefont {T.}~\bibnamefont {Cui}},\ }\bibfield  {title} {\bibinfo {title} {{Synthesis and Superconductivity of Ternary A15--\ch{(Lu,Y)4H23} at High Pressures}},\ }\href@noop {} {\bibfield  {journal} {\bibinfo  {journal} {J. Am. Chem. Soc.}\ } (\bibinfo {year} {2025})}\BibitemShut {NoStop}%
\bibitem [{\citenamefont {Du}\ \emph {et~al.}(2024)\citenamefont {Du}, \citenamefont {Huang}, \citenamefont {Zhang}, \citenamefont {Wang}, \citenamefont {Song}, \citenamefont {Duan},\ and\ \citenamefont {Cui}}]{du2024high}%
  \BibitemOpen
  \bibfield  {author} {\bibinfo {author} {\bibfnamefont {M.}~\bibnamefont {Du}}, \bibinfo {author} {\bibfnamefont {H.}~\bibnamefont {Huang}}, \bibinfo {author} {\bibfnamefont {Z.}~\bibnamefont {Zhang}}, \bibinfo {author} {\bibfnamefont {M.}~\bibnamefont {Wang}}, \bibinfo {author} {\bibfnamefont {H.}~\bibnamefont {Song}}, \bibinfo {author} {\bibfnamefont {D.}~\bibnamefont {Duan}},\ and\ \bibinfo {author} {\bibfnamefont {T.}~\bibnamefont {Cui}},\ }\bibfield  {title} {\bibinfo {title} {{High-Temperature Superconductivity in Perovskite Hydride Below \SI{10}{GPa}}},\ }\href@noop {} {\bibfield  {journal} {\bibinfo  {journal} {Adv. Sci.}\ }\textbf {\bibinfo {volume} {11}},\ \bibinfo {pages} {2408370} (\bibinfo {year} {2024})}\BibitemShut {NoStop}%
\bibitem [{\citenamefont {Jiang}\ \emph {et~al.}(2024)\citenamefont {Jiang}, \citenamefont {Zhang}, \citenamefont {Song}, \citenamefont {Ma}, \citenamefont {Sun}, \citenamefont {Miao}, \citenamefont {Cui},\ and\ \citenamefont {Duan}}]{jiang2024ternary}%
  \BibitemOpen
  \bibfield  {author} {\bibinfo {author} {\bibfnamefont {Q.}~\bibnamefont {Jiang}}, \bibinfo {author} {\bibfnamefont {Z.}~\bibnamefont {Zhang}}, \bibinfo {author} {\bibfnamefont {H.}~\bibnamefont {Song}}, \bibinfo {author} {\bibfnamefont {Y.}~\bibnamefont {Ma}}, \bibinfo {author} {\bibfnamefont {Y.}~\bibnamefont {Sun}}, \bibinfo {author} {\bibfnamefont {M.}~\bibnamefont {Miao}}, \bibinfo {author} {\bibfnamefont {T.}~\bibnamefont {Cui}},\ and\ \bibinfo {author} {\bibfnamefont {D.}~\bibnamefont {Duan}},\ }\bibfield  {title} {\bibinfo {title} {{Ternary superconducting hydrides stabilized via Th and Ce elements at mild pressures}},\ }\href@noop {} {\bibfield  {journal} {\bibinfo  {journal} {Fundam. Res.}\ }\textbf {\bibinfo {volume} {4}},\ \bibinfo {pages} {550} (\bibinfo {year} {2024})}\BibitemShut {NoStop}%
\bibitem [{\citenamefont {He}\ \emph {et~al.}(2024)\citenamefont {He}, \citenamefont {Zhao}, \citenamefont {Xie}, \citenamefont {Hermann}, \citenamefont {Hemley}, \citenamefont {Liu},\ and\ \citenamefont {Ma}}]{he2024predicted}%
  \BibitemOpen
  \bibfield  {author} {\bibinfo {author} {\bibfnamefont {X.-L.}\ \bibnamefont {He}}, \bibinfo {author} {\bibfnamefont {W.}~\bibnamefont {Zhao}}, \bibinfo {author} {\bibfnamefont {Y.}~\bibnamefont {Xie}}, \bibinfo {author} {\bibfnamefont {A.}~\bibnamefont {Hermann}}, \bibinfo {author} {\bibfnamefont {R.~J.}\ \bibnamefont {Hemley}}, \bibinfo {author} {\bibfnamefont {H.}~\bibnamefont {Liu}},\ and\ \bibinfo {author} {\bibfnamefont {Y.}~\bibnamefont {Ma}},\ }\bibfield  {title} {\bibinfo {title} {{Predicted hot superconductivity in \ch{LaSc2H24} under pressure}},\ }\href@noop {} {\bibfield  {journal} {\bibinfo  {journal} {Proc. Natl. Acad. Sci. U.S.A.}\ }\textbf {\bibinfo {volume} {121}},\ \bibinfo {pages} {e2401840121} (\bibinfo {year} {2024})}\BibitemShut {NoStop}%
\bibitem [{\citenamefont {Lucrezi}\ \emph {et~al.}(2023)\citenamefont {Lucrezi}, \citenamefont {Kogler}, \citenamefont {Di~Cataldo}, \citenamefont {Aichhorn}, \citenamefont {Boeri},\ and\ \citenamefont {Heil}}]{lucrezi2023quantum}%
  \BibitemOpen
  \bibfield  {author} {\bibinfo {author} {\bibfnamefont {R.}~\bibnamefont {Lucrezi}}, \bibinfo {author} {\bibfnamefont {E.}~\bibnamefont {Kogler}}, \bibinfo {author} {\bibfnamefont {S.}~\bibnamefont {Di~Cataldo}}, \bibinfo {author} {\bibfnamefont {M.}~\bibnamefont {Aichhorn}}, \bibinfo {author} {\bibfnamefont {L.}~\bibnamefont {Boeri}},\ and\ \bibinfo {author} {\bibfnamefont {C.}~\bibnamefont {Heil}},\ }\bibfield  {title} {\bibinfo {title} {{Quantum lattice dynamics and their importance in ternary superhydride clathrates}},\ }\href@noop {} {\bibfield  {journal} {\bibinfo  {journal} {Commun. Phys.}\ }\textbf {\bibinfo {volume} {6}},\ \bibinfo {pages} {298} (\bibinfo {year} {2023})}\BibitemShut {NoStop}%
\bibitem [{\citenamefont {Zhang}\ \emph {et~al.}(2022{\natexlab{b}})\citenamefont {Zhang}, \citenamefont {Cui}, \citenamefont {Hutcheon}, \citenamefont {Shipley}, \citenamefont {Song}, \citenamefont {Du}, \citenamefont {Kresin}, \citenamefont {Duan}, \citenamefont {Pickard},\ and\ \citenamefont {Yao}}]{zhang2022design}%
  \BibitemOpen
  \bibfield  {author} {\bibinfo {author} {\bibfnamefont {Z.}~\bibnamefont {Zhang}}, \bibinfo {author} {\bibfnamefont {T.}~\bibnamefont {Cui}}, \bibinfo {author} {\bibfnamefont {M.~J.}\ \bibnamefont {Hutcheon}}, \bibinfo {author} {\bibfnamefont {A.~M.}\ \bibnamefont {Shipley}}, \bibinfo {author} {\bibfnamefont {H.}~\bibnamefont {Song}}, \bibinfo {author} {\bibfnamefont {M.}~\bibnamefont {Du}}, \bibinfo {author} {\bibfnamefont {V.~Z.}\ \bibnamefont {Kresin}}, \bibinfo {author} {\bibfnamefont {D.}~\bibnamefont {Duan}}, \bibinfo {author} {\bibfnamefont {C.~J.}\ \bibnamefont {Pickard}},\ and\ \bibinfo {author} {\bibfnamefont {Y.}~\bibnamefont {Yao}},\ }\bibfield  {title} {\bibinfo {title} {{Design principles for high-temperature superconductors with a hydrogen-based alloy backbone at moderate pressure}},\ }\href@noop {} {\bibfield  {journal} {\bibinfo  {journal} {Phys. Rev. Lett.}\ }\textbf {\bibinfo {volume} {128}},\ \bibinfo {pages} {047001} (\bibinfo {year} {2022}{\natexlab{b}})}\BibitemShut {NoStop}%
\bibitem [{\citenamefont {Zhao}\ \emph {et~al.}(2022)\citenamefont {Zhao}, \citenamefont {Zhang}, \citenamefont {Li}, \citenamefont {Ding}, \citenamefont {Liu},\ and\ \citenamefont {Yang}}]{Zhao2022emergent}%
  \BibitemOpen
  \bibfield  {author} {\bibinfo {author} {\bibfnamefont {Y.}~\bibnamefont {Zhao}}, \bibinfo {author} {\bibfnamefont {X.}~\bibnamefont {Zhang}}, \bibinfo {author} {\bibfnamefont {X.}~\bibnamefont {Li}}, \bibinfo {author} {\bibfnamefont {S.}~\bibnamefont {Ding}}, \bibinfo {author} {\bibfnamefont {Y.}~\bibnamefont {Liu}},\ and\ \bibinfo {author} {\bibfnamefont {G.}~\bibnamefont {Yang}},\ }\bibfield  {title} {\bibinfo {title} {{Emergent superconductivity in \ch{K2ReH9} under pressure}},\ }\href@noop {} {\bibfield  {journal} {\bibinfo  {journal} {J. Mater. Chem. C}\ }\textbf {\bibinfo {volume} {10}},\ \bibinfo {pages} {14626} (\bibinfo {year} {2022})}\BibitemShut {NoStop}%
\bibitem [{\citenamefont {Gao}\ \emph {et~al.}(2021{\natexlab{b}})\citenamefont {Gao}, \citenamefont {Yan}, \citenamefont {Lu},\ and\ \citenamefont {Xiang}}]{gao2021phonon}%
  \BibitemOpen
  \bibfield  {author} {\bibinfo {author} {\bibfnamefont {M.}~\bibnamefont {Gao}}, \bibinfo {author} {\bibfnamefont {X.-W.}\ \bibnamefont {Yan}}, \bibinfo {author} {\bibfnamefont {Z.-Y.}\ \bibnamefont {Lu}},\ and\ \bibinfo {author} {\bibfnamefont {T.}~\bibnamefont {Xiang}},\ }\bibfield  {title} {\bibinfo {title} {{Phonon-mediated high-temperature superconductivity in the ternary borohydride \ch{KB2H8} under pressure near \SI{12}{GPa}}},\ }\href@noop {} {\bibfield  {journal} {\bibinfo  {journal} {Phys. Rev. B}\ }\textbf {\bibinfo {volume} {104}},\ \bibinfo {pages} {L100504} (\bibinfo {year} {2021}{\natexlab{b}})}\BibitemShut {NoStop}%
\bibitem [{\citenamefont {Sun}\ \emph {et~al.}(2019)\citenamefont {Sun}, \citenamefont {Lv}, \citenamefont {Xie}, \citenamefont {Liu},\ and\ \citenamefont {Ma}}]{sun2019route}%
  \BibitemOpen
  \bibfield  {author} {\bibinfo {author} {\bibfnamefont {Y.}~\bibnamefont {Sun}}, \bibinfo {author} {\bibfnamefont {J.}~\bibnamefont {Lv}}, \bibinfo {author} {\bibfnamefont {Y.}~\bibnamefont {Xie}}, \bibinfo {author} {\bibfnamefont {H.}~\bibnamefont {Liu}},\ and\ \bibinfo {author} {\bibfnamefont {Y.}~\bibnamefont {Ma}},\ }\bibfield  {title} {\bibinfo {title} {{Route to a superconducting phase above room temperature in electron-doped hydride compounds under high pressure}},\ }\href@noop {} {\bibfield  {journal} {\bibinfo  {journal} {Phys. Rev. Lett.}\ }\textbf {\bibinfo {volume} {123}},\ \bibinfo {pages} {097001} (\bibinfo {year} {2019})}\BibitemShut {NoStop}%
\bibitem [{\citenamefont {Song}\ \emph {et~al.}(2023)\citenamefont {Song}, \citenamefont {Bi}, \citenamefont {Nakamoto}, \citenamefont {Shimizu}, \citenamefont {Liu}, \citenamefont {Zou}, \citenamefont {Liu}, \citenamefont {Wang},\ and\ \citenamefont {Ma}}]{song2023stoichiometric}%
  \BibitemOpen
  \bibfield  {author} {\bibinfo {author} {\bibfnamefont {Y.}~\bibnamefont {Song}}, \bibinfo {author} {\bibfnamefont {J.}~\bibnamefont {Bi}}, \bibinfo {author} {\bibfnamefont {Y.}~\bibnamefont {Nakamoto}}, \bibinfo {author} {\bibfnamefont {K.}~\bibnamefont {Shimizu}}, \bibinfo {author} {\bibfnamefont {H.}~\bibnamefont {Liu}}, \bibinfo {author} {\bibfnamefont {B.}~\bibnamefont {Zou}}, \bibinfo {author} {\bibfnamefont {G.}~\bibnamefont {Liu}}, \bibinfo {author} {\bibfnamefont {H.}~\bibnamefont {Wang}},\ and\ \bibinfo {author} {\bibfnamefont {Y.}~\bibnamefont {Ma}},\ }\bibfield  {title} {\bibinfo {title} {{Stoichiometric ternary superhydride \ch{LaBeH8} as a new template for high-temperature superconductivity at \SI{110}{K} under \SI{80}{GPa}}},\ }\href@noop {} {\bibfield  {journal} {\bibinfo  {journal} {Phys. Rev. Lett.}\ }\textbf {\bibinfo {volume} {130}},\ \bibinfo {pages} {266001} (\bibinfo {year} {2023})}\BibitemShut {NoStop}%
\bibitem [{\citenamefont {Song}\ \emph {et~al.}(2024)\citenamefont {Song}, \citenamefont {Hao}, \citenamefont {Wei}, \citenamefont {He}, \citenamefont {Liu}, \citenamefont {Ma}, \citenamefont {Liu}, \citenamefont {Wang}, \citenamefont {Niu}, \citenamefont {Wang} \emph {et~al.}}]{song2024superconductivity}%
  \BibitemOpen
  \bibfield  {author} {\bibinfo {author} {\bibfnamefont {X.}~\bibnamefont {Song}}, \bibinfo {author} {\bibfnamefont {X.}~\bibnamefont {Hao}}, \bibinfo {author} {\bibfnamefont {X.}~\bibnamefont {Wei}}, \bibinfo {author} {\bibfnamefont {X.-L.}\ \bibnamefont {He}}, \bibinfo {author} {\bibfnamefont {H.}~\bibnamefont {Liu}}, \bibinfo {author} {\bibfnamefont {L.}~\bibnamefont {Ma}}, \bibinfo {author} {\bibfnamefont {G.}~\bibnamefont {Liu}}, \bibinfo {author} {\bibfnamefont {H.}~\bibnamefont {Wang}}, \bibinfo {author} {\bibfnamefont {J.}~\bibnamefont {Niu}}, \bibinfo {author} {\bibfnamefont {S.}~\bibnamefont {Wang}}, \emph {et~al.},\ }\bibfield  {title} {\bibinfo {title} {{Superconductivity above 105 K in nonclathrate ternary lanthanum borohydride below megabar pressure}},\ }\href@noop {} {\bibfield  {journal} {\bibinfo  {journal} {J. Am. Chem. Soc.}\ }\textbf {\bibinfo {volume} {146}},\ \bibinfo {pages} {13797} (\bibinfo {year} {2024})}\BibitemShut {NoStop}%
\bibitem [{\citenamefont {Yvon}\ and\ \citenamefont {Renaudin}(2006)}]{Yvon2006}%
  \BibitemOpen
  \bibfield  {author} {\bibinfo {author} {\bibfnamefont {K.}~\bibnamefont {Yvon}}\ and\ \bibinfo {author} {\bibfnamefont {G.}~\bibnamefont {Renaudin}},\ }\bibinfo {title} {{Hydrides: solid state transition metal complexes}},\ in\ \href@noop {} {\emph {\bibinfo {booktitle} {{Encyclopedia of Inorganic and Bioinorganic Chemistry}}}}\ (\bibinfo  {publisher} {John Wiley \& Sons Ltd},\ \bibinfo {year} {2006})\ p.\ \bibinfo {pages} {1814–1846}\BibitemShut {NoStop}%
\bibitem [{\citenamefont {Bronger}\ \emph {et~al.}(1999)\citenamefont {Bronger}, \citenamefont {Brassard}, \citenamefont {M{\"u}ller}, \citenamefont {Lebech},\ and\ \citenamefont {Schultz}}]{bronger1999k2reh9}%
  \BibitemOpen
  \bibfield  {author} {\bibinfo {author} {\bibfnamefont {W.}~\bibnamefont {Bronger}}, \bibinfo {author} {\bibfnamefont {L.~{\`a}.}\ \bibnamefont {Brassard}}, \bibinfo {author} {\bibfnamefont {P.}~\bibnamefont {M{\"u}ller}}, \bibinfo {author} {\bibfnamefont {B.}~\bibnamefont {Lebech}},\ and\ \bibinfo {author} {\bibfnamefont {T.}~\bibnamefont {Schultz}},\ }\bibfield  {title} {\bibinfo {title} {{\ch{K2ReH9}, eine Neubestimmung der Struktur}},\ }\href@noop {} {\bibfield  {journal} {\bibinfo  {journal} {Z. Anorg. Allg. Chem.}\ }\textbf {\bibinfo {volume} {625}},\ \bibinfo {pages} {1143} (\bibinfo {year} {1999})}\BibitemShut {NoStop}%
\bibitem [{\citenamefont {Bronger}\ \emph {et~al.}(2002)\citenamefont {Bronger}, \citenamefont {Sommer}, \citenamefont {Auffermann},\ and\ \citenamefont {M{\"u}ller}}]{bronger2002new}%
  \BibitemOpen
  \bibfield  {author} {\bibinfo {author} {\bibfnamefont {W.}~\bibnamefont {Bronger}}, \bibinfo {author} {\bibfnamefont {T.}~\bibnamefont {Sommer}}, \bibinfo {author} {\bibfnamefont {G.}~\bibnamefont {Auffermann}},\ and\ \bibinfo {author} {\bibfnamefont {P.}~\bibnamefont {M{\"u}ller}},\ }\bibfield  {title} {\bibinfo {title} {{New alkali metal osmium-and ruthenium hydrides}},\ }\href@noop {} {\bibfield  {journal} {\bibinfo  {journal} {J. Alloys Compd.}\ }\textbf {\bibinfo {volume} {330}},\ \bibinfo {pages} {536} (\bibinfo {year} {2002})}\BibitemShut {NoStop}%
\bibitem [{\citenamefont {Orgaz}(2000)}]{orgaz2000electronic}%
  \BibitemOpen
  \bibfield  {author} {\bibinfo {author} {\bibfnamefont {E.}~\bibnamefont {Orgaz}},\ }\bibfield  {title} {\bibinfo {title} {{Electronic structure of magnetic ternary alkali metal--manganese hydrides}},\ }\href@noop {} {\bibfield  {journal} {\bibinfo  {journal} {Phys. Rev. B}\ }\textbf {\bibinfo {volume} {61}},\ \bibinfo {pages} {7989} (\bibinfo {year} {2000})}\BibitemShut {NoStop}%
\bibitem [{\citenamefont {Ivanov}\ \emph {et~al.}(1989)\citenamefont {Ivanov}, \citenamefont {Konstanchuk}, \citenamefont {Stepanov}, \citenamefont {Jie}, \citenamefont {Pezat},\ and\ \citenamefont {Darriet}}]{ivanov1989ternary}%
  \BibitemOpen
  \bibfield  {author} {\bibinfo {author} {\bibfnamefont {E.~Y.}\ \bibnamefont {Ivanov}}, \bibinfo {author} {\bibfnamefont {I.}~\bibnamefont {Konstanchuk}}, \bibinfo {author} {\bibfnamefont {A.}~\bibnamefont {Stepanov}}, \bibinfo {author} {\bibfnamefont {Y.}~\bibnamefont {Jie}}, \bibinfo {author} {\bibfnamefont {M.}~\bibnamefont {Pezat}},\ and\ \bibinfo {author} {\bibfnamefont {B.}~\bibnamefont {Darriet}},\ }\bibfield  {title} {\bibinfo {title} {{The ternary system magnesium-cobalt-hydrogen}},\ }\href@noop {} {\bibfield  {journal} {\bibinfo  {journal} {Inorg. Chem.}\ }\textbf {\bibinfo {volume} {28}},\ \bibinfo {pages} {613} (\bibinfo {year} {1989})}\BibitemShut {NoStop}%
\bibitem [{\citenamefont {Olofsson-M{\aa}rtensson}\ \emph {et~al.}(2000)\citenamefont {Olofsson-M{\aa}rtensson}, \citenamefont {H{\"a}ussermann}, \citenamefont {Tomkinson},\ and\ \citenamefont {Nor{\'e}us}}]{olofsson2000stabilization}%
  \BibitemOpen
  \bibfield  {author} {\bibinfo {author} {\bibfnamefont {M.}~\bibnamefont {Olofsson-M{\aa}rtensson}}, \bibinfo {author} {\bibfnamefont {U.}~\bibnamefont {H{\"a}ussermann}}, \bibinfo {author} {\bibfnamefont {J.}~\bibnamefont {Tomkinson}},\ and\ \bibinfo {author} {\bibfnamefont {D.}~\bibnamefont {Nor{\'e}us}},\ }\bibfield  {title} {\bibinfo {title} {{Stabilization of electron-dense {Palladium- Hydrido} complexes in solid-state hydrides}},\ }\href@noop {} {\bibfield  {journal} {\bibinfo  {journal} {J. Am. Chem. Soc.}\ }\textbf {\bibinfo {volume} {122}},\ \bibinfo {pages} {6960} (\bibinfo {year} {2000})}\BibitemShut {NoStop}%
\bibitem [{\citenamefont {Olofsson}\ \emph {et~al.}(1998)\citenamefont {Olofsson}, \citenamefont {Kritikos},\ and\ \citenamefont {Nor{\'e}us}}]{olofsson1998first}%
  \BibitemOpen
  \bibfield  {author} {\bibinfo {author} {\bibfnamefont {M.}~\bibnamefont {Olofsson}}, \bibinfo {author} {\bibfnamefont {M.}~\bibnamefont {Kritikos}},\ and\ \bibinfo {author} {\bibfnamefont {D.}~\bibnamefont {Nor{\'e}us}},\ }\bibfield  {title} {\bibinfo {title} {{The First Trigonal Planar Transition Metal- Hydrogen Complex in \ch{NaBaPdH3}}},\ }\href@noop {} {\bibfield  {journal} {\bibinfo  {journal} {Inorg. Chem.}\ }\textbf {\bibinfo {volume} {37}},\ \bibinfo {pages} {2900} (\bibinfo {year} {1998})}\BibitemShut {NoStop}%
\bibitem [{\citenamefont {Bonhomme}\ \emph {et~al.}(1992)\citenamefont {Bonhomme}, \citenamefont {Yvon},\ and\ \citenamefont {Fischer}}]{bonhomme1992tetragonal}%
  \BibitemOpen
  \bibfield  {author} {\bibinfo {author} {\bibfnamefont {F.}~\bibnamefont {Bonhomme}}, \bibinfo {author} {\bibfnamefont {K.}~\bibnamefont {Yvon}},\ and\ \bibinfo {author} {\bibfnamefont {P.}~\bibnamefont {Fischer}},\ }\bibfield  {title} {\bibinfo {title} {{Tetragonal trimagnesium ruthenium trideuteride, \ch{Mg3RuD3}, containing dinuclear \ch{[Ru2D6]^12-} complex anions}},\ }\href@noop {} {\bibfield  {journal} {\bibinfo  {journal} {J. Alloys Compd.}\ }\textbf {\bibinfo {volume} {186}},\ \bibinfo {pages} {309} (\bibinfo {year} {1992})}\BibitemShut {NoStop}%
\bibitem [{\citenamefont {Nor{\'e}us}\ \emph {et~al.}(1988)\citenamefont {Nor{\'e}us}, \citenamefont {T{\"o}rnroos}, \citenamefont {B{\"o}rje}, \citenamefont {Szabo}, \citenamefont {Bronger}, \citenamefont {Spittank}, \citenamefont {Auffermann},\ and\ \citenamefont {M{\"u}ller}}]{noreus1988na2pdh2}%
  \BibitemOpen
  \bibfield  {author} {\bibinfo {author} {\bibfnamefont {D.}~\bibnamefont {Nor{\'e}us}}, \bibinfo {author} {\bibfnamefont {K.}~\bibnamefont {T{\"o}rnroos}}, \bibinfo {author} {\bibfnamefont {A.}~\bibnamefont {B{\"o}rje}}, \bibinfo {author} {\bibfnamefont {T.}~\bibnamefont {Szabo}}, \bibinfo {author} {\bibfnamefont {W.}~\bibnamefont {Bronger}}, \bibinfo {author} {\bibfnamefont {H.}~\bibnamefont {Spittank}}, \bibinfo {author} {\bibfnamefont {G.}~\bibnamefont {Auffermann}},\ and\ \bibinfo {author} {\bibfnamefont {P.}~\bibnamefont {M{\"u}ller}},\ }\bibfield  {title} {\bibinfo {title} {{\ch{Na2PdH2}, a hydride with a novel linear \ch{[PdH2]} complex}},\ }\href@noop {} {\bibfield  {journal} {\bibinfo  {journal} {J. Less Common Met.}\ }\textbf {\bibinfo {volume} {139}},\ \bibinfo {pages} {233} (\bibinfo {year} {1988})}\BibitemShut {NoStop}%
\bibitem [{\citenamefont {Muramatsu}\ \emph {et~al.}(2015)\citenamefont {Muramatsu}, \citenamefont {Wanene}, \citenamefont {Somayazulu}, \citenamefont {Vinitsky}, \citenamefont {Chandra}, \citenamefont {Strobel}, \citenamefont {Struzhkin},\ and\ \citenamefont {Hemley}}]{muramatsu2015metallization}%
  \BibitemOpen
  \bibfield  {author} {\bibinfo {author} {\bibfnamefont {T.}~\bibnamefont {Muramatsu}}, \bibinfo {author} {\bibfnamefont {W.~K.}\ \bibnamefont {Wanene}}, \bibinfo {author} {\bibfnamefont {M.}~\bibnamefont {Somayazulu}}, \bibinfo {author} {\bibfnamefont {E.}~\bibnamefont {Vinitsky}}, \bibinfo {author} {\bibfnamefont {D.}~\bibnamefont {Chandra}}, \bibinfo {author} {\bibfnamefont {T.~A.}\ \bibnamefont {Strobel}}, \bibinfo {author} {\bibfnamefont {V.~V.}\ \bibnamefont {Struzhkin}},\ and\ \bibinfo {author} {\bibfnamefont {R.~J.}\ \bibnamefont {Hemley}},\ }\bibfield  {title} {\bibinfo {title} {{Metallization and superconductivity in the hydrogen-rich ionic salt \ch{BaReH9}}},\ }\href@noop {} {\bibfield  {journal} {\bibinfo  {journal} {J. Phys. Chem. C}\ }\textbf {\bibinfo {volume} {119}},\ \bibinfo {pages} {18007} (\bibinfo {year} {2015})}\BibitemShut {NoStop}%
\bibitem [{\citenamefont {Meng}\ \emph {et~al.}(2019)\citenamefont {Meng}, \citenamefont {Sakata}, \citenamefont {Shimizu}, \citenamefont {Iijima}, \citenamefont {Saitoh}, \citenamefont {Sato}, \citenamefont {Takagi},\ and\ \citenamefont {Orimo}}]{meng2019superconductivity}%
  \BibitemOpen
  \bibfield  {author} {\bibinfo {author} {\bibfnamefont {D.}~\bibnamefont {Meng}}, \bibinfo {author} {\bibfnamefont {M.}~\bibnamefont {Sakata}}, \bibinfo {author} {\bibfnamefont {K.}~\bibnamefont {Shimizu}}, \bibinfo {author} {\bibfnamefont {Y.}~\bibnamefont {Iijima}}, \bibinfo {author} {\bibfnamefont {H.}~\bibnamefont {Saitoh}}, \bibinfo {author} {\bibfnamefont {T.}~\bibnamefont {Sato}}, \bibinfo {author} {\bibfnamefont {S.}~\bibnamefont {Takagi}},\ and\ \bibinfo {author} {\bibfnamefont {S.-i.}\ \bibnamefont {Orimo}},\ }\bibfield  {title} {\bibinfo {title} {{Superconductivity of the hydrogen-rich metal hydride \ch{Li5MoH11} under high pressure}},\ }\href@noop {} {\bibfield  {journal} {\bibinfo  {journal} {Phys. Rev. B}\ }\textbf {\bibinfo {volume} {99}},\ \bibinfo {pages} {024508} (\bibinfo {year} {2019})}\BibitemShut {NoStop}%
\bibitem [{\citenamefont {Sanna}\ \emph {et~al.}(2024)\citenamefont {Sanna}, \citenamefont {Cerqueira}, \citenamefont {Fang}, \citenamefont {Errea}, \citenamefont {Ludwig},\ and\ \citenamefont {Marques}}]{sanna2024prediction}%
  \BibitemOpen
  \bibfield  {author} {\bibinfo {author} {\bibfnamefont {A.}~\bibnamefont {Sanna}}, \bibinfo {author} {\bibfnamefont {T.~F.}\ \bibnamefont {Cerqueira}}, \bibinfo {author} {\bibfnamefont {Y.-W.}\ \bibnamefont {Fang}}, \bibinfo {author} {\bibfnamefont {I.}~\bibnamefont {Errea}}, \bibinfo {author} {\bibfnamefont {A.}~\bibnamefont {Ludwig}},\ and\ \bibinfo {author} {\bibfnamefont {M.~A.}\ \bibnamefont {Marques}},\ }\bibfield  {title} {\bibinfo {title} {{Prediction of ambient pressure conventional superconductivity above \SI{80}{K} in hydride compounds}},\ }\href@noop {} {\bibfield  {journal} {\bibinfo  {journal} {npj Comput. Mater.}\ }\textbf {\bibinfo {volume} {10}},\ \bibinfo {pages} {44} (\bibinfo {year} {2024})}\BibitemShut {NoStop}%
\bibitem [{\citenamefont {Zheng}\ \emph {et~al.}(2024)\citenamefont {Zheng}, \citenamefont {Zhang}, \citenamefont {Wu}, \citenamefont {Wu}, \citenamefont {Lin}, \citenamefont {Wang}, \citenamefont {Fang}, \citenamefont {Wang}, \citenamefont {Antropov}, \citenamefont {Sun} \emph {et~al.}}]{zheng2024prediction}%
  \BibitemOpen
  \bibfield  {author} {\bibinfo {author} {\bibfnamefont {F.}~\bibnamefont {Zheng}}, \bibinfo {author} {\bibfnamefont {Z.}~\bibnamefont {Zhang}}, \bibinfo {author} {\bibfnamefont {Z.}~\bibnamefont {Wu}}, \bibinfo {author} {\bibfnamefont {S.}~\bibnamefont {Wu}}, \bibinfo {author} {\bibfnamefont {Q.}~\bibnamefont {Lin}}, \bibinfo {author} {\bibfnamefont {R.}~\bibnamefont {Wang}}, \bibinfo {author} {\bibfnamefont {Y.}~\bibnamefont {Fang}}, \bibinfo {author} {\bibfnamefont {C.-Z.}\ \bibnamefont {Wang}}, \bibinfo {author} {\bibfnamefont {V.}~\bibnamefont {Antropov}}, \bibinfo {author} {\bibfnamefont {Y.}~\bibnamefont {Sun}}, \emph {et~al.},\ }\bibfield  {title} {\bibinfo {title} {{Prediction of ambient pressure superconductivity in cubic ternary hydrides with MH6 octahedra}},\ }\href@noop {} {\bibfield  {journal} {\bibinfo  {journal} {Mater. Today Phys.}\ }\textbf {\bibinfo {volume} {42}},\ \bibinfo {pages} {101374} (\bibinfo {year} {2024})}\BibitemShut {NoStop}%
\bibitem [{\citenamefont {Dolui}\ \emph {et~al.}(2024)\citenamefont {Dolui}, \citenamefont {Conway}, \citenamefont {Heil}, \citenamefont {Strobel}, \citenamefont {Prasankumar},\ and\ \citenamefont {Pickard}}]{dolui2024feasible}%
  \BibitemOpen
  \bibfield  {author} {\bibinfo {author} {\bibfnamefont {K.}~\bibnamefont {Dolui}}, \bibinfo {author} {\bibfnamefont {L.~J.}\ \bibnamefont {Conway}}, \bibinfo {author} {\bibfnamefont {C.}~\bibnamefont {Heil}}, \bibinfo {author} {\bibfnamefont {T.~A.}\ \bibnamefont {Strobel}}, \bibinfo {author} {\bibfnamefont {R.~P.}\ \bibnamefont {Prasankumar}},\ and\ \bibinfo {author} {\bibfnamefont {C.~J.}\ \bibnamefont {Pickard}},\ }\bibfield  {title} {\bibinfo {title} {{Feasible route to high-temperature ambient-pressure hydride superconductivity}},\ }\href@noop {} {\bibfield  {journal} {\bibinfo  {journal} {Phys. Rev. Lett.}\ }\textbf {\bibinfo {volume} {132}},\ \bibinfo {pages} {166001} (\bibinfo {year} {2024})}\BibitemShut {NoStop}%
\bibitem [{\citenamefont {Hansen}\ \emph {et~al.}(2024)\citenamefont {Hansen}, \citenamefont {Conway}, \citenamefont {Dolui}, \citenamefont {Heil}, \citenamefont {Pickard}, \citenamefont {Pakhomova}, \citenamefont {Mezouar}, \citenamefont {Kunz}, \citenamefont {Prasankumar},\ and\ \citenamefont {Strobel}}]{hansen2024}%
  \BibitemOpen
  \bibfield  {author} {\bibinfo {author} {\bibfnamefont {M.~F.}\ \bibnamefont {Hansen}}, \bibinfo {author} {\bibfnamefont {L.~J.}\ \bibnamefont {Conway}}, \bibinfo {author} {\bibfnamefont {K.}~\bibnamefont {Dolui}}, \bibinfo {author} {\bibfnamefont {C.}~\bibnamefont {Heil}}, \bibinfo {author} {\bibfnamefont {C.~J.}\ \bibnamefont {Pickard}}, \bibinfo {author} {\bibfnamefont {A.}~\bibnamefont {Pakhomova}}, \bibinfo {author} {\bibfnamefont {M.}~\bibnamefont {Mezouar}}, \bibinfo {author} {\bibfnamefont {M.}~\bibnamefont {Kunz}}, \bibinfo {author} {\bibfnamefont {R.~P.}\ \bibnamefont {Prasankumar}},\ and\ \bibinfo {author} {\bibfnamefont {T.~A.}\ \bibnamefont {Strobel}},\ }\bibfield  {title} {\bibinfo {title} {Synthesis of \ch{Mg2IrH5}: A potential pathway to high-${T}_{c}$ hydride superconductivity at ambient pressure},\ }\href {https://doi.org/10.1103/PhysRevB.110.214513} {\bibfield  {journal} {\bibinfo  {journal} {Phys. Rev. B}\ }\textbf {\bibinfo {volume} {110}},\ \bibinfo {pages} {214513} (\bibinfo {year}
  {2024})}\BibitemShut {NoStop}%
\bibitem [{\citenamefont {Reilly}\ \emph {et~al.}(2016)\citenamefont {Reilly}, \citenamefont {Cooper}, \citenamefont {Adjiman}, \citenamefont {Bhattacharya}, \citenamefont {Boese}, \citenamefont {Brandenburg}, \citenamefont {Bygrave}, \citenamefont {Bylsma}, \citenamefont {Campbell}, \citenamefont {Car} \emph {et~al.}}]{reilly2016report}%
  \BibitemOpen
  \bibfield  {author} {\bibinfo {author} {\bibfnamefont {A.~M.}\ \bibnamefont {Reilly}}, \bibinfo {author} {\bibfnamefont {R.~I.}\ \bibnamefont {Cooper}}, \bibinfo {author} {\bibfnamefont {C.~S.}\ \bibnamefont {Adjiman}}, \bibinfo {author} {\bibfnamefont {S.}~\bibnamefont {Bhattacharya}}, \bibinfo {author} {\bibfnamefont {A.~D.}\ \bibnamefont {Boese}}, \bibinfo {author} {\bibfnamefont {J.~G.}\ \bibnamefont {Brandenburg}}, \bibinfo {author} {\bibfnamefont {P.~J.}\ \bibnamefont {Bygrave}}, \bibinfo {author} {\bibfnamefont {R.}~\bibnamefont {Bylsma}}, \bibinfo {author} {\bibfnamefont {J.~E.}\ \bibnamefont {Campbell}}, \bibinfo {author} {\bibfnamefont {R.}~\bibnamefont {Car}}, \emph {et~al.},\ }\bibfield  {title} {\bibinfo {title} {{Report on the sixth blind test of organic crystal structure prediction methods}},\ }\href@noop {} {\bibfield  {journal} {\bibinfo  {journal} {Acta Crystallogr. B}\ }\textbf {\bibinfo {volume} {72}},\ \bibinfo {pages} {439} (\bibinfo {year} {2016})}\BibitemShut {NoStop}%
\bibitem [{\citenamefont {Saal}\ \emph {et~al.}(2013)\citenamefont {Saal}, \citenamefont {Kirklin}, \citenamefont {Aykol},\ and\ \citenamefont {Meredig}}]{saal2013materials}%
  \BibitemOpen
  \bibfield  {author} {\bibinfo {author} {\bibfnamefont {J.~E.}\ \bibnamefont {Saal}}, \bibinfo {author} {\bibfnamefont {S.}~\bibnamefont {Kirklin}}, \bibinfo {author} {\bibfnamefont {M.}~\bibnamefont {Aykol}},\ and\ \bibinfo {author} {\bibfnamefont {B.}~\bibnamefont {Meredig}},\ }\bibfield  {title} {\bibinfo {title} {{Materials design and discovery with high-throughput density functional theory: the open quantum materials database (OQMD)}},\ }\href@noop {} {\bibfield  {journal} {\bibinfo  {journal} {JOM}\ }\textbf {\bibinfo {volume} {65}},\ \bibinfo {pages} {1501} (\bibinfo {year} {2013})}\BibitemShut {NoStop}%
\bibitem [{\citenamefont {Schmidt}\ \emph {et~al.}(2023)\citenamefont {Schmidt}, \citenamefont {Hoffmann}, \citenamefont {Wang}, \citenamefont {Borlido}, \citenamefont {Carri{\c{c}}o}, \citenamefont {Cerqueira}, \citenamefont {Botti},\ and\ \citenamefont {Marques}}]{schmidt2023machine}%
  \BibitemOpen
  \bibfield  {author} {\bibinfo {author} {\bibfnamefont {J.}~\bibnamefont {Schmidt}}, \bibinfo {author} {\bibfnamefont {N.}~\bibnamefont {Hoffmann}}, \bibinfo {author} {\bibfnamefont {H.-C.}\ \bibnamefont {Wang}}, \bibinfo {author} {\bibfnamefont {P.}~\bibnamefont {Borlido}}, \bibinfo {author} {\bibfnamefont {P.~J.}\ \bibnamefont {Carri{\c{c}}o}}, \bibinfo {author} {\bibfnamefont {T.~F.}\ \bibnamefont {Cerqueira}}, \bibinfo {author} {\bibfnamefont {S.}~\bibnamefont {Botti}},\ and\ \bibinfo {author} {\bibfnamefont {M.~A.}\ \bibnamefont {Marques}},\ }\bibfield  {title} {\bibinfo {title} {{Machine-Learning-Assisted Determination of the Global Zero-Temperature Phase Diagram of Materials}},\ }\href@noop {} {\bibfield  {journal} {\bibinfo  {journal} {Adv. Mater.}\ }\textbf {\bibinfo {volume} {35}},\ \bibinfo {pages} {2210788} (\bibinfo {year} {2023})}\BibitemShut {NoStop}%
\bibitem [{\citenamefont {Jain}\ \emph {et~al.}(2013)\citenamefont {Jain}, \citenamefont {Ong}, \citenamefont {Hautier}, \citenamefont {Chen}, \citenamefont {Richards}, \citenamefont {Dacek}, \citenamefont {Cholia}, \citenamefont {Gunter}, \citenamefont {Skinner}, \citenamefont {Ceder} \emph {et~al.}}]{jain2013commentary}%
  \BibitemOpen
  \bibfield  {author} {\bibinfo {author} {\bibfnamefont {A.}~\bibnamefont {Jain}}, \bibinfo {author} {\bibfnamefont {S.~P.}\ \bibnamefont {Ong}}, \bibinfo {author} {\bibfnamefont {G.}~\bibnamefont {Hautier}}, \bibinfo {author} {\bibfnamefont {W.}~\bibnamefont {Chen}}, \bibinfo {author} {\bibfnamefont {W.~D.}\ \bibnamefont {Richards}}, \bibinfo {author} {\bibfnamefont {S.}~\bibnamefont {Dacek}}, \bibinfo {author} {\bibfnamefont {S.}~\bibnamefont {Cholia}}, \bibinfo {author} {\bibfnamefont {D.}~\bibnamefont {Gunter}}, \bibinfo {author} {\bibfnamefont {D.}~\bibnamefont {Skinner}}, \bibinfo {author} {\bibfnamefont {G.}~\bibnamefont {Ceder}}, \emph {et~al.},\ }\bibfield  {title} {\bibinfo {title} {{Commentary: The Materials Project: A materials genome approach to accelerating materials innovation}},\ }\href@noop {} {\bibfield  {journal} {\bibinfo  {journal} {APL Mater.}\ }\textbf {\bibinfo {volume} {1}} (\bibinfo {year} {2013})}\BibitemShut {NoStop}%
\bibitem [{\citenamefont {Nayeb-Hashemi}\ and\ \citenamefont {Clark}(1985)}]{nayeb1985mg}%
  \BibitemOpen
  \bibfield  {author} {\bibinfo {author} {\bibfnamefont {A.}~\bibnamefont {Nayeb-Hashemi}}\ and\ \bibinfo {author} {\bibfnamefont {J.~B.}\ \bibnamefont {Clark}},\ }\bibfield  {title} {\bibinfo {title} {{The Mg-Pt (Magnesium-Platinum) System}},\ }\href@noop {} {\bibfield  {journal} {\bibinfo  {journal} {Bull. Alloy Phase Diagr.}\ }\textbf {\bibinfo {volume} {6}},\ \bibinfo {pages} {533} (\bibinfo {year} {1985})}\BibitemShut {NoStop}%
\bibitem [{\citenamefont {Ellinger}\ \emph {et~al.}(1955)\citenamefont {Ellinger}, \citenamefont {Holley~Jr}, \citenamefont {McInteer}, \citenamefont {Pavone}, \citenamefont {Potter}, \citenamefont {Staritzky},\ and\ \citenamefont {Zachariasen}}]{ellinger1955preparation}%
  \BibitemOpen
  \bibfield  {author} {\bibinfo {author} {\bibfnamefont {F.}~\bibnamefont {Ellinger}}, \bibinfo {author} {\bibfnamefont {C.}~\bibnamefont {Holley~Jr}}, \bibinfo {author} {\bibfnamefont {B.}~\bibnamefont {McInteer}}, \bibinfo {author} {\bibfnamefont {D.}~\bibnamefont {Pavone}}, \bibinfo {author} {\bibfnamefont {R.}~\bibnamefont {Potter}}, \bibinfo {author} {\bibfnamefont {E.}~\bibnamefont {Staritzky}},\ and\ \bibinfo {author} {\bibfnamefont {W.}~\bibnamefont {Zachariasen}},\ }\bibfield  {title} {\bibinfo {title} {{The preparation and some properties of magnesium Hydride}},\ }\href@noop {} {\bibfield  {journal} {\bibinfo  {journal} {Journal of the American Chemical Society}\ }\textbf {\bibinfo {volume} {77}},\ \bibinfo {pages} {2647} (\bibinfo {year} {1955})}\BibitemShut {NoStop}%
\bibitem [{\citenamefont {Ferro}\ and\ \citenamefont {Rambaldi}(1960)}]{ferro1960research}%
  \BibitemOpen
  \bibfield  {author} {\bibinfo {author} {\bibfnamefont {R.}~\bibnamefont {Ferro}}\ and\ \bibinfo {author} {\bibfnamefont {G.}~\bibnamefont {Rambaldi}},\ }\bibfield  {title} {\bibinfo {title} {{Research on the alloys of noble metals with the more electropositive elements: III. Micrographic and X-ray examination of some magnesium-platinum alloys}},\ }\href@noop {} {\bibfield  {journal} {\bibinfo  {journal} {Journal of the Less Common Metals}\ }\textbf {\bibinfo {volume} {2}},\ \bibinfo {pages} {383} (\bibinfo {year} {1960})}\BibitemShut {NoStop}%
\bibitem [{\citenamefont {Evans}\ \emph {et~al.}(2024)\citenamefont {Evans}, \citenamefont {Bergsma}, \citenamefont {Merkys}, \citenamefont {Andersen}, \citenamefont {Andersson}, \citenamefont {Beltr{\'a}n}, \citenamefont {Blokhin}, \citenamefont {Boland}, \citenamefont {Balderas}, \citenamefont {Choudhary} \emph {et~al.}}]{evans2024developments}%
  \BibitemOpen
  \bibfield  {author} {\bibinfo {author} {\bibfnamefont {M.~L.}\ \bibnamefont {Evans}}, \bibinfo {author} {\bibfnamefont {J.}~\bibnamefont {Bergsma}}, \bibinfo {author} {\bibfnamefont {A.}~\bibnamefont {Merkys}}, \bibinfo {author} {\bibfnamefont {C.~W.}\ \bibnamefont {Andersen}}, \bibinfo {author} {\bibfnamefont {O.~B.}\ \bibnamefont {Andersson}}, \bibinfo {author} {\bibfnamefont {D.}~\bibnamefont {Beltr{\'a}n}}, \bibinfo {author} {\bibfnamefont {E.}~\bibnamefont {Blokhin}}, \bibinfo {author} {\bibfnamefont {T.~M.}\ \bibnamefont {Boland}}, \bibinfo {author} {\bibfnamefont {R.~C.}\ \bibnamefont {Balderas}}, \bibinfo {author} {\bibfnamefont {K.}~\bibnamefont {Choudhary}}, \emph {et~al.},\ }\bibfield  {title} {\bibinfo {title} {{Developments and applications of the OPTIMADE API for materials discovery, design, and data exchange}},\ }\href@noop {} {\bibfield  {journal} {\bibinfo  {journal} {Digit. Discov.}\ }\textbf {\bibinfo {volume} {3}},\ \bibinfo {pages} {1509} (\bibinfo {year} {2024})}\BibitemShut {NoStop}%
\bibitem [{\citenamefont {Pickard}(2022)}]{pickard2022ephemeral}%
  \BibitemOpen
  \bibfield  {author} {\bibinfo {author} {\bibfnamefont {C.~J.}\ \bibnamefont {Pickard}},\ }\bibfield  {title} {\bibinfo {title} {{Ephemeral data derived potentials for random structure search}},\ }\href@noop {} {\bibfield  {journal} {\bibinfo  {journal} {Phys. Rev. B}\ }\textbf {\bibinfo {volume} {106}},\ \bibinfo {pages} {014102} (\bibinfo {year} {2022})}\BibitemShut {NoStop}%
\bibitem [{\citenamefont {Pickard}\ and\ \citenamefont {Needs}(2011)}]{pickard2011ab}%
  \BibitemOpen
  \bibfield  {author} {\bibinfo {author} {\bibfnamefont {C.~J.}\ \bibnamefont {Pickard}}\ and\ \bibinfo {author} {\bibfnamefont {R.}~\bibnamefont {Needs}},\ }\bibfield  {title} {\bibinfo {title} {{Ab initio random structure searching}},\ }\href@noop {} {\bibfield  {journal} {\bibinfo  {journal} {J. Phys.: Condens. Matter}\ }\textbf {\bibinfo {volume} {23}},\ \bibinfo {pages} {053201} (\bibinfo {year} {2011})}\BibitemShut {NoStop}%
\bibitem [{\citenamefont {Giannozzi}\ \emph {et~al.}(2009)\citenamefont {Giannozzi}, \citenamefont {Baroni}, \citenamefont {Bonini}, \citenamefont {Calandra}, \citenamefont {Car}, \citenamefont {Cavazzoni}, \citenamefont {Ceresoli}, \citenamefont {Chiarotti}, \citenamefont {Cococcioni}, \citenamefont {Dabo} \emph {et~al.}}]{giannozzi2009quantum}%
  \BibitemOpen
  \bibfield  {author} {\bibinfo {author} {\bibfnamefont {P.}~\bibnamefont {Giannozzi}}, \bibinfo {author} {\bibfnamefont {S.}~\bibnamefont {Baroni}}, \bibinfo {author} {\bibfnamefont {N.}~\bibnamefont {Bonini}}, \bibinfo {author} {\bibfnamefont {M.}~\bibnamefont {Calandra}}, \bibinfo {author} {\bibfnamefont {R.}~\bibnamefont {Car}}, \bibinfo {author} {\bibfnamefont {C.}~\bibnamefont {Cavazzoni}}, \bibinfo {author} {\bibfnamefont {D.}~\bibnamefont {Ceresoli}}, \bibinfo {author} {\bibfnamefont {G.~L.}\ \bibnamefont {Chiarotti}}, \bibinfo {author} {\bibfnamefont {M.}~\bibnamefont {Cococcioni}}, \bibinfo {author} {\bibfnamefont {I.}~\bibnamefont {Dabo}}, \emph {et~al.},\ }\bibfield  {title} {\bibinfo {title} {{QUANTUM ESPRESSO: a modular and open-source software project for quantum simulations of materials}},\ }\href@noop {} {\bibfield  {journal} {\bibinfo  {journal} {J. Phys.: Condens. Matter}\ }\textbf {\bibinfo {volume} {21}},\ \bibinfo {pages} {395502} (\bibinfo {year} {2009})}\BibitemShut {NoStop}%
\bibitem [{\citenamefont {Perdew}\ \emph {et~al.}(1996)\citenamefont {Perdew}, \citenamefont {Burke},\ and\ \citenamefont {Ernzerhof}}]{perdew1996generalized}%
  \BibitemOpen
  \bibfield  {author} {\bibinfo {author} {\bibfnamefont {J.~P.}\ \bibnamefont {Perdew}}, \bibinfo {author} {\bibfnamefont {K.}~\bibnamefont {Burke}},\ and\ \bibinfo {author} {\bibfnamefont {M.}~\bibnamefont {Ernzerhof}},\ }\bibfield  {title} {\bibinfo {title} {{Generalized gradient approximation made simple}},\ }\href@noop {} {\bibfield  {journal} {\bibinfo  {journal} {Phys. Rev. Lett.}\ }\textbf {\bibinfo {volume} {77}},\ \bibinfo {pages} {3865} (\bibinfo {year} {1996})}\BibitemShut {NoStop}%
\bibitem [{\citenamefont {Garrity}\ \emph {et~al.}(2014)\citenamefont {Garrity}, \citenamefont {Bennett}, \citenamefont {Rabe},\ and\ \citenamefont {Vanderbilt}}]{garrity2014pseudopotentials}%
  \BibitemOpen
  \bibfield  {author} {\bibinfo {author} {\bibfnamefont {K.~F.}\ \bibnamefont {Garrity}}, \bibinfo {author} {\bibfnamefont {J.~W.}\ \bibnamefont {Bennett}}, \bibinfo {author} {\bibfnamefont {K.~M.}\ \bibnamefont {Rabe}},\ and\ \bibinfo {author} {\bibfnamefont {D.}~\bibnamefont {Vanderbilt}},\ }\bibfield  {title} {\bibinfo {title} {{Pseudopotentials for high-throughput DFT calculations}},\ }\href@noop {} {\bibfield  {journal} {\bibinfo  {journal} {Comput. Mater. Sci.}\ }\textbf {\bibinfo {volume} {81}},\ \bibinfo {pages} {446} (\bibinfo {year} {2014})}\BibitemShut {NoStop}%
\bibitem [{\citenamefont {Monkhorst}\ and\ \citenamefont {Pack}(1976)}]{monkhorst1976special}%
  \BibitemOpen
  \bibfield  {author} {\bibinfo {author} {\bibfnamefont {H.~J.}\ \bibnamefont {Monkhorst}}\ and\ \bibinfo {author} {\bibfnamefont {J.~D.}\ \bibnamefont {Pack}},\ }\bibfield  {title} {\bibinfo {title} {{Special points for Brillouin-zone integrations}},\ }\href@noop {} {\bibfield  {journal} {\bibinfo  {journal} {Phys. Rev. B}\ }\textbf {\bibinfo {volume} {13}},\ \bibinfo {pages} {5188} (\bibinfo {year} {1976})}\BibitemShut {NoStop}%
\bibitem [{\citenamefont {Togo}\ \emph {et~al.}(2008)\citenamefont {Togo}, \citenamefont {Oba},\ and\ \citenamefont {Tanaka}}]{togo2008first}%
  \BibitemOpen
  \bibfield  {author} {\bibinfo {author} {\bibfnamefont {A.}~\bibnamefont {Togo}}, \bibinfo {author} {\bibfnamefont {F.}~\bibnamefont {Oba}},\ and\ \bibinfo {author} {\bibfnamefont {I.}~\bibnamefont {Tanaka}},\ }\bibfield  {title} {\bibinfo {title} {{First-principles calculations of the ferroelastic transition between rutile-type and \ch{CaCl2}-type \ch{SiO2} at high pressures}},\ }\href@noop {} {\bibfield  {journal} {\bibinfo  {journal} {Phys. Rev. B}\ }\textbf {\bibinfo {volume} {78}},\ \bibinfo {pages} {134106} (\bibinfo {year} {2008})}\BibitemShut {NoStop}%
\bibitem [{\citenamefont {Akahama}\ and\ \citenamefont {Kawamura}(2010)}]{Akahama_2010}%
  \BibitemOpen
  \bibfield  {author} {\bibinfo {author} {\bibfnamefont {Y.}~\bibnamefont {Akahama}}\ and\ \bibinfo {author} {\bibfnamefont {H.}~\bibnamefont {Kawamura}},\ }\bibfield  {title} {\bibinfo {title} {Pressure calibration of diamond anvil raman gauge to 410 gpa},\ }\href@noop {} {\bibfield  {journal} {\bibinfo  {journal} {J. Phys.: Conf. Ser.}\ }\textbf {\bibinfo {volume} {215}},\ \bibinfo {pages} {012195} (\bibinfo {year} {2010})}\BibitemShut {NoStop}%
\bibitem [{\citenamefont {Anderson}\ \emph {et~al.}(1989)\citenamefont {Anderson}, \citenamefont {Isaak},\ and\ \citenamefont {Yamamoto}}]{anderson1989anharmonicity}%
  \BibitemOpen
  \bibfield  {author} {\bibinfo {author} {\bibfnamefont {O.~L.}\ \bibnamefont {Anderson}}, \bibinfo {author} {\bibfnamefont {D.~G.}\ \bibnamefont {Isaak}},\ and\ \bibinfo {author} {\bibfnamefont {S.}~\bibnamefont {Yamamoto}},\ }\bibfield  {title} {\bibinfo {title} {{Anharmonicity and the equation of state for gold}},\ }\href@noop {} {\bibfield  {journal} {\bibinfo  {journal} {J. Appl. Phys.}\ }\textbf {\bibinfo {volume} {65}},\ \bibinfo {pages} {1534} (\bibinfo {year} {1989})}\BibitemShut {NoStop}%
\bibitem [{\citenamefont {Holmes}\ \emph {et~al.}(1989)\citenamefont {Holmes}, \citenamefont {Moriarty}, \citenamefont {Gathers},\ and\ \citenamefont {Nellis}}]{holmes1989equation}%
  \BibitemOpen
  \bibfield  {author} {\bibinfo {author} {\bibfnamefont {N.}~\bibnamefont {Holmes}}, \bibinfo {author} {\bibfnamefont {J.}~\bibnamefont {Moriarty}}, \bibinfo {author} {\bibfnamefont {G.}~\bibnamefont {Gathers}},\ and\ \bibinfo {author} {\bibfnamefont {W.}~\bibnamefont {Nellis}},\ }\bibfield  {title} {\bibinfo {title} {{The equation of state of platinum to \SI{660}{GPa} (\SI{6.6}{Mbar})}},\ }\href@noop {} {\bibfield  {journal} {\bibinfo  {journal} {J. Appl. Phys.}\ }\textbf {\bibinfo {volume} {66}},\ \bibinfo {pages} {2962} (\bibinfo {year} {1989})}\BibitemShut {NoStop}%
\bibitem [{\citenamefont {Heinz}\ and\ \citenamefont {Jeanloz}(1987)}]{Heinz1987}%
  \BibitemOpen
  \bibfield  {author} {\bibinfo {author} {\bibfnamefont {D.~L.}\ \bibnamefont {Heinz}}\ and\ \bibinfo {author} {\bibfnamefont {R.}~\bibnamefont {Jeanloz}},\ }\bibinfo {title} {Temperature measurements in the laser-heated diamond cell},\ in\ \href@noop {} {\emph {\bibinfo {booktitle} {High‐Pressure Research in Mineral Physics: A Volume in Honor of Syun‐iti Akimoto}}}\ (\bibinfo  {publisher} {American Geophysical Union (AGU)},\ \bibinfo {year} {1987})\ pp.\ \bibinfo {pages} {113--127}\BibitemShut {NoStop}%
\bibitem [{\citenamefont {Mezouar}\ \emph {et~al.}(2024)\citenamefont {Mezouar}, \citenamefont {Garbarino}, \citenamefont {Bauchau}, \citenamefont {Morgenroth}, \citenamefont {Martel}, \citenamefont {Petitdemange}, \citenamefont {Got}, \citenamefont {Clavel}, \citenamefont {Moyne}, \citenamefont {Van Der~Kleij} \emph {et~al.}}]{mezouar2024high}%
  \BibitemOpen
  \bibfield  {author} {\bibinfo {author} {\bibfnamefont {M.}~\bibnamefont {Mezouar}}, \bibinfo {author} {\bibfnamefont {G.}~\bibnamefont {Garbarino}}, \bibinfo {author} {\bibfnamefont {S.}~\bibnamefont {Bauchau}}, \bibinfo {author} {\bibfnamefont {W.}~\bibnamefont {Morgenroth}}, \bibinfo {author} {\bibfnamefont {K.}~\bibnamefont {Martel}}, \bibinfo {author} {\bibfnamefont {S.}~\bibnamefont {Petitdemange}}, \bibinfo {author} {\bibfnamefont {P.}~\bibnamefont {Got}}, \bibinfo {author} {\bibfnamefont {C.}~\bibnamefont {Clavel}}, \bibinfo {author} {\bibfnamefont {A.}~\bibnamefont {Moyne}}, \bibinfo {author} {\bibfnamefont {H.-P.}\ \bibnamefont {Van Der~Kleij}}, \emph {et~al.},\ }\bibfield  {title} {\bibinfo {title} {{The high flux nano-X-ray diffraction, fluorescence and imaging beamline ID27 for science under extreme conditions on the ESRF Extremely Brilliant Source}},\ }\href {https://doi.org/https://doi.org/10.1080/08957959.2024.2363932} {\bibfield  {journal} {\bibinfo  {journal} {High Pressure Research}\
  }\textbf {\bibinfo {volume} {44}},\ \bibinfo {pages} {171} (\bibinfo {year} {2024})}\BibitemShut {NoStop}%
\bibitem [{\citenamefont {Prescher}\ and\ \citenamefont {Prakapenka}(2015)}]{prescher2015dioptas}%
  \BibitemOpen
  \bibfield  {author} {\bibinfo {author} {\bibfnamefont {C.}~\bibnamefont {Prescher}}\ and\ \bibinfo {author} {\bibfnamefont {V.~B.}\ \bibnamefont {Prakapenka}},\ }\bibfield  {title} {\bibinfo {title} {{DIOPTAS: a program for reduction of two-dimensional X-ray diffraction data and data exploration}},\ }\href@noop {} {\bibfield  {journal} {\bibinfo  {journal} {High Press. Res.}\ }\textbf {\bibinfo {volume} {35}},\ \bibinfo {pages} {223} (\bibinfo {year} {2015})}\BibitemShut {NoStop}%
\bibitem [{\citenamefont {Toby}\ and\ \citenamefont {Von~Dreele}(2013)}]{toby2013gsas}%
  \BibitemOpen
  \bibfield  {author} {\bibinfo {author} {\bibfnamefont {B.~H.}\ \bibnamefont {Toby}}\ and\ \bibinfo {author} {\bibfnamefont {R.~B.}\ \bibnamefont {Von~Dreele}},\ }\bibfield  {title} {\bibinfo {title} {{GSAS-II: the genesis of a modern open-source all purpose crystallography software package}},\ }\href@noop {} {\bibfield  {journal} {\bibinfo  {journal} {J. Appl. Crystallogr.}\ }\textbf {\bibinfo {volume} {46}},\ \bibinfo {pages} {544} (\bibinfo {year} {2013})}\BibitemShut {NoStop}%
\bibitem [{\citenamefont {Gonzalez-Platas}\ \emph {et~al.}(2016)\citenamefont {Gonzalez-Platas}, \citenamefont {Alvaro}, \citenamefont {Nestola},\ and\ \citenamefont {Angel}}]{gonzalez2016eosfit7}%
  \BibitemOpen
  \bibfield  {author} {\bibinfo {author} {\bibfnamefont {J.}~\bibnamefont {Gonzalez-Platas}}, \bibinfo {author} {\bibfnamefont {M.}~\bibnamefont {Alvaro}}, \bibinfo {author} {\bibfnamefont {F.}~\bibnamefont {Nestola}},\ and\ \bibinfo {author} {\bibfnamefont {R.}~\bibnamefont {Angel}},\ }\bibfield  {title} {\bibinfo {title} {{EosFit7-GUI: a new graphical user interface for equation of state calculations, analyses and teaching}},\ }\href@noop {} {\bibfield  {journal} {\bibinfo  {journal} {J. Appl. Crystallogr.}\ }\textbf {\bibinfo {volume} {49}},\ \bibinfo {pages} {1377} (\bibinfo {year} {2016})}\BibitemShut {NoStop}%
\bibitem [{\citenamefont {Bronger}\ and\ \citenamefont {Breil}(1998)}]{bronger1998calcium}%
  \BibitemOpen
  \bibfield  {author} {\bibinfo {author} {\bibfnamefont {W.}~\bibnamefont {Bronger}}\ and\ \bibinfo {author} {\bibfnamefont {L.}~\bibnamefont {Breil}},\ }\bibfield  {title} {\bibinfo {title} {{Calcium--Rhodium--Hydride---Synthese und Struktur}},\ }\href@noop {} {\bibfield  {journal} {\bibinfo  {journal} {Z. Anorg. Allg. Chem.}\ }\textbf {\bibinfo {volume} {624}},\ \bibinfo {pages} {1819} (\bibinfo {year} {1998})}\BibitemShut {NoStop}%
\bibitem [{sup()}]{supp}%
  \BibitemOpen
  \href@noop {} {}\bibinfo {note} {See Supplemental Material at [URL-will-be-inserted-by-publisher] for supporting figures and tables.}\BibitemShut {Stop}%
\bibitem [{\citenamefont {Staroverov}\ \emph {et~al.}(2004)\citenamefont {Staroverov}, \citenamefont {Scuseria}, \citenamefont {Tao},\ and\ \citenamefont {Perdew}}]{staroverov2004tests}%
  \BibitemOpen
  \bibfield  {author} {\bibinfo {author} {\bibfnamefont {V.~N.}\ \bibnamefont {Staroverov}}, \bibinfo {author} {\bibfnamefont {G.~E.}\ \bibnamefont {Scuseria}}, \bibinfo {author} {\bibfnamefont {J.}~\bibnamefont {Tao}},\ and\ \bibinfo {author} {\bibfnamefont {J.~P.}\ \bibnamefont {Perdew}},\ }\bibfield  {title} {\bibinfo {title} {{Tests of a ladder of density functionals for bulk solids and surfaces}},\ }\href@noop {} {\bibfield  {journal} {\bibinfo  {journal} {Phys. Rev. B}\ }\textbf {\bibinfo {volume} {69}},\ \bibinfo {pages} {075102} (\bibinfo {year} {2004})}\BibitemShut {NoStop}%
\bibitem [{\citenamefont {Wu}\ and\ \citenamefont {Cohen}(2006)}]{wu2006more}%
  \BibitemOpen
  \bibfield  {author} {\bibinfo {author} {\bibfnamefont {Z.}~\bibnamefont {Wu}}\ and\ \bibinfo {author} {\bibfnamefont {R.~E.}\ \bibnamefont {Cohen}},\ }\bibfield  {title} {\bibinfo {title} {{More accurate generalized gradient approximation for solids}},\ }\href@noop {} {\bibfield  {journal} {\bibinfo  {journal} {Phys. Rev. B}\ }\textbf {\bibinfo {volume} {73}},\ \bibinfo {pages} {235116} (\bibinfo {year} {2006})}\BibitemShut {NoStop}%
\bibitem [{\citenamefont {Bader}(1991)}]{bader1991quantum}%
  \BibitemOpen
  \bibfield  {author} {\bibinfo {author} {\bibfnamefont {R.~F.}\ \bibnamefont {Bader}},\ }\bibfield  {title} {\bibinfo {title} {{A quantum theory of molecular structure and its applications}},\ }\href@noop {} {\bibfield  {journal} {\bibinfo  {journal} {Chem. Rev.}\ }\textbf {\bibinfo {volume} {91}},\ \bibinfo {pages} {893} (\bibinfo {year} {1991})}\BibitemShut {NoStop}%
\bibitem [{\citenamefont {Firman}\ and\ \citenamefont {Landis}(1998)}]{firman1998structure}%
  \BibitemOpen
  \bibfield  {author} {\bibinfo {author} {\bibfnamefont {T.~K.}\ \bibnamefont {Firman}}\ and\ \bibinfo {author} {\bibfnamefont {C.~R.}\ \bibnamefont {Landis}},\ }\bibfield  {title} {\bibinfo {title} {{Structure and electron counting in ternary transition metal hydrides}},\ }\href@noop {} {\bibfield  {journal} {\bibinfo  {journal} {J. Am. Chem. Soc.}\ }\textbf {\bibinfo {volume} {120}},\ \bibinfo {pages} {12650} (\bibinfo {year} {1998})}\BibitemShut {NoStop}%
\bibitem [{\citenamefont {Kadir}\ and\ \citenamefont {Nor{\'e}us}(1993)}]{kadir1993a2h2}%
  \BibitemOpen
  \bibfield  {author} {\bibinfo {author} {\bibfnamefont {K.}~\bibnamefont {Kadir}}\ and\ \bibinfo {author} {\bibfnamefont {D.}~\bibnamefont {Nor{\'e}us}},\ }\bibfield  {title} {\bibinfo {title} {{\ch{A2H2[PtH4]} (A = Sr and Ba), two Hydrides with a Layered Structure Type where \ch{[Pt(II)H4]^2-} Complexes and Hydrogen Atoms in Tetrahedral Interstices Share the same Alkaline Earth Counter Ions}},\ }\href@noop {} {\bibfield  {journal} {\bibinfo  {journal} {Z. Phys. Chem.}\ }\textbf {\bibinfo {volume} {179}},\ \bibinfo {pages} {237} (\bibinfo {year} {1993})}\BibitemShut {NoStop}%
\bibitem [{\citenamefont {Bronger}\ and\ \citenamefont {Brassard}(1996)}]{bronger1996li2pth2}%
  \BibitemOpen
  \bibfield  {author} {\bibinfo {author} {\bibfnamefont {W.}~\bibnamefont {Bronger}}\ and\ \bibinfo {author} {\bibfnamefont {L.}~\bibnamefont {Brassard}},\ }\bibfield  {title} {\bibinfo {title} {{\ch{Li2PtH2}, Darstellung und Struktur}},\ }\href@noop {} {\bibfield  {journal} {\bibinfo  {journal} {Z. Anorg. Allg. Chem.}\ }\textbf {\bibinfo {volume} {622}},\ \bibinfo {pages} {462} (\bibinfo {year} {1996})}\BibitemShut {NoStop}%
\bibitem [{\citenamefont {Bronger}\ \emph {et~al.}(1984)\citenamefont {Bronger}, \citenamefont {M{\"u}ller}, \citenamefont {Schmitz},\ and\ \citenamefont {Spittank}}]{bronger1984synthese}%
  \BibitemOpen
  \bibfield  {author} {\bibinfo {author} {\bibfnamefont {W.}~\bibnamefont {Bronger}}, \bibinfo {author} {\bibfnamefont {P.}~\bibnamefont {M{\"u}ller}}, \bibinfo {author} {\bibfnamefont {D.}~\bibnamefont {Schmitz}},\ and\ \bibinfo {author} {\bibfnamefont {H.}~\bibnamefont {Spittank}},\ }\bibfield  {title} {\bibinfo {title} {{Synthese und Struktur von \ch{Na2PtH4}, einem tern{\"a}ren Hydrid mit quadratisch planaren \ch{[PtH4]^2-}-Baugruppen}},\ }\href@noop {} {\bibfield  {journal} {\bibinfo  {journal} {Z. Anorg. Allg. Chem.}\ }\textbf {\bibinfo {volume} {516}},\ \bibinfo {pages} {35} (\bibinfo {year} {1984})}\BibitemShut {NoStop}%
\bibitem [{\citenamefont {Bronger}\ \emph {et~al.}(1988)\citenamefont {Bronger}, \citenamefont {Auffermann},\ and\ \citenamefont {M\"uller}}]{bronger1988darstellung}%
  \BibitemOpen
  \bibfield  {author} {\bibinfo {author} {\bibfnamefont {W.}~\bibnamefont {Bronger}}, \bibinfo {author} {\bibfnamefont {G.}~\bibnamefont {Auffermann}},\ and\ \bibinfo {author} {\bibfnamefont {P.}~\bibnamefont {M\"uller}},\ }\bibfield  {title} {\bibinfo {title} {{Darstellung und Struktur tern\"arer Platinhydride \ch{A3PtH5} mit A$\equiv$ K, Rb oder Cs}},\ }\href@noop {} {\bibfield  {journal} {\bibinfo  {journal} {Z. Anorg. Allg. Chem.}\ }\textbf {\bibinfo {volume} {566}},\ \bibinfo {pages} {31} (\bibinfo {year} {1988})}\BibitemShut {NoStop}%
\bibitem [{\citenamefont {Bronger}\ and\ \citenamefont {Brassard}(1995)}]{bronger1995li5pt2h9}%
  \BibitemOpen
  \bibfield  {author} {\bibinfo {author} {\bibfnamefont {W.}~\bibnamefont {Bronger}}\ and\ \bibinfo {author} {\bibfnamefont {L.~{\`a}.}\ \bibnamefont {Brassard}},\ }\bibfield  {title} {\bibinfo {title} {{\ch{Li5Pt2H9}, a Complex Hydride Containing Isolated \ch{[Pt2H9]^5-} Ions}},\ }\href@noop {} {\bibfield  {journal} {\bibinfo  {journal} {Angew. Chem. Int. Ed. Engl.}\ }\textbf {\bibinfo {volume} {34}},\ \bibinfo {pages} {898} (\bibinfo {year} {1995})}\BibitemShut {NoStop}%
\bibitem [{\citenamefont {Bronger}\ and\ \citenamefont {Auffermann}(1995)}]{bronger1995high}%
  \BibitemOpen
  \bibfield  {author} {\bibinfo {author} {\bibfnamefont {W.}~\bibnamefont {Bronger}}\ and\ \bibinfo {author} {\bibfnamefont {G.}~\bibnamefont {Auffermann}},\ }\bibfield  {title} {\bibinfo {title} {{High pressure synthesis and structure of \ch{Na2PtH6}}},\ }\href@noop {} {\bibfield  {journal} {\bibinfo  {journal} {J. Alloys Compd.}\ }\textbf {\bibinfo {volume} {219}},\ \bibinfo {pages} {45} (\bibinfo {year} {1995})}\BibitemShut {NoStop}%
\bibitem [{\citenamefont {Ginzburg}\ and\ \citenamefont {Landau}(2009)}]{ginzburg2009theory}%
  \BibitemOpen
  \bibfield  {author} {\bibinfo {author} {\bibfnamefont {V.~L.}\ \bibnamefont {Ginzburg}}\ and\ \bibinfo {author} {\bibfnamefont {L.~D.}\ \bibnamefont {Landau}},\ }\bibinfo {title} {{On the Theory of Superconductivity}},\ in\ \href@noop {} {\emph {\bibinfo {booktitle} {On Superconductivity and Superfluidity: A Scientific Autobiography}}}\ (\bibinfo  {publisher} {Springer Berlin Heidelberg},\ \bibinfo {year} {2009})\ pp.\ \bibinfo {pages} {113--137}\BibitemShut {NoStop}%
\bibitem [{\citenamefont {Allen}\ and\ \citenamefont {Dynes}(1975)}]{allen1975transition}%
  \BibitemOpen
  \bibfield  {author} {\bibinfo {author} {\bibfnamefont {P.~B.}\ \bibnamefont {Allen}}\ and\ \bibinfo {author} {\bibfnamefont {R.}~\bibnamefont {Dynes}},\ }\bibfield  {title} {\bibinfo {title} {Transition temperature of strong-coupled superconductors reanalyzed},\ }\href@noop {} {\bibfield  {journal} {\bibinfo  {journal} {Phys. Rev. B}\ }\textbf {\bibinfo {volume} {12}},\ \bibinfo {pages} {905} (\bibinfo {year} {1975})}\BibitemShut {NoStop}%
\bibitem [{\citenamefont {Dunning}\ \emph {et~al.}(2026)\citenamefont {Dunning}, \citenamefont {Hansen}, \citenamefont {Hübner},\ and\ \citenamefont {Strobel}}]{2023_CH-6822_ESRF}%
  \BibitemOpen
  \bibfield  {author} {\bibinfo {author} {\bibfnamefont {S.}~\bibnamefont {Dunning}}, \bibinfo {author} {\bibfnamefont {M.~F.}\ \bibnamefont {Hansen}}, \bibinfo {author} {\bibfnamefont {J.-M.}\ \bibnamefont {Hübner}},\ and\ \bibinfo {author} {\bibfnamefont {T.}~\bibnamefont {Strobel}},\ }\href {https://doi.org/10.15151/ESRF-ES-1345038308} {\bibinfo {title} {{New Boron-Stabilized Carbon Clathrate Structures [{ESRF} dataset]}}} (\bibinfo {year} {2026})\BibitemShut {NoStop}%
\end{thebibliography}%

\end{document}